\documentclass[11pt]{article}
\usepackage{float}
\usepackage{color}
\usepackage{mymacros}
\usepackage{chngcntr}
\usepackage{verbatim}
\usepackage{amsthm}
\usepackage{graphics}
\usepackage{graphicx}
\usepackage{xcolor}
\usepackage{amsmath}
\usepackage{amssymb}

\usepackage{mathrsfs}
\usepackage{bbold}
\usepackage{cases}
\usepackage{epsfig}
\usepackage{epstopdf}
\usepackage{hyperref}

\title{\boldmath Chaos and multifold complexity for an inverted harmonic oscillator}

\author{Le-Chen Qu,$^{a,b}$ \footnote{qulch20@lzu.edu.cn}}
\author{Hong-Yue Jiang$^{a,b}$ \footnote{jianghy21@lzu.edu.cn}}
\author{and Yu-Xiao Liu$^{a,b}$
\footnote{liuyx@lzu.edu.cn, corresponding author}}

\affiliation{
$^{a}$Lanzhou Center for Theoretical Physics, Key Laboratory of Theoretical Physics of Gansu Province, School of Physical Science and Technology, Lanzhou University, Lanzhou 730000, China\\
$^{b}$Institute of Theoretical Physics and Research Center of Gravitation, Lanzhou University, Lanzhou 730000, China\\
}

\abstract{We examine the multifold complexity and Loschmidt echo for an inverted harmonic oscillator. We give analytic expressions for any number of precursors, implementing multiple backward and forward time evolutions of the quantum state, at the leading order in the perturbation. We prove that complexity is dominated by the longest permutation of the given time combination in an alternating ``zig-zag'' order, the exact same result obtained with holography.
We conjecture that the general structure for multifold complexity should hold true universally for generic quantum systems, in the limit of a large number of precursors.}

\begin{document}
\maketitle
\section{Introduction}
The precise definition of quantum chaos still remains an open question. Unlike classical chaos, quantum chaos is only partially understood \cite{{1},{Jahnke:2018off}}. It is believed that proper characterization of quantum chaos will help us understand thermalization, transport properties in quantum many-body systems, and black hole information loss, among others. Consequently, to make progress in these areas it seem crucial to deepen our understanding quantum chaos at a fundamental level.

Due to the non-linear nature of the equation of motion, the evolution of classical systems may be particularly sensitive to the beginning conditions, which accounts for the appearance of chaos.  The deviation of nearby trajectories in phase space is measured by the Poisson bracket of position ${q}(t)$ and the momentum ${p}(0)$. For chaotic systems, the expectation is that nearby trajectories will deviate exponentially as the system evolves,
\begin{equation}\label{eqPBchaos}
\left\{{q}(t), {p} (0)\right\}=\frac{\partial {q}(t)}{\partial {p} (0)}\sim e^{\lambda t},
\end{equation}
where $\lambda$ is the so-called Lyapunov exponent \cite{F. Haake}. After this exponential growth regime, the position of the particle in phase space become unpredictable, the system is said to have lost memory of its initial conditions, a phenomenon dubbed the \emph{butterfly effect}.

Quantum systems, however, are more subtle than their classical counterparts. On one hand, their evolution is dictated by Schr\"odinger's equation, which is linear in time. On the other hand, both position and momentum are upgraded to operators $\hat{q}$ and $\hat{p}$, and are no longer useful to describe their phase space. A natural generalization of (\ref{eqPBchaos}) would be to upgrade the Poisson bracket to a commutator, and take an expectation value in the corresponding quantum state, $\left\{{q}(t), {p} (0)\right\}\to\langle[\hat{q}(t), \hat{p} (0)]\rangle$. However, as chaos ought to be a characteristic of the system, one would be better off thinking about a typical state, e.g. drawn from an ensemble, rather than considering specific microstates. The canonical choice is to consider a thermal ensemble, where the expectation value $\langle\cdot\rangle$ is replaced by the thermal expectation value, $\langle\cdot\rangle_\beta=Z^{-1}\tr[e^{-\beta H}\cdot]$. There is still one more issue, and it is the fact that the two-point functions used to calculate the aforementioned quantity thermalize quickly, in a time scale of order of the inverse temperature $t_\text{th}\sim\beta$. Within this time window there is just not enough time to develop an exponential growth akin to (\ref{eqPBchaos}). We understand this as as result of cancelations between positive and negative contributions to the two-point function, which are generic after the system reaches local equilibration. One simple fix is to consider
the square of the commutator bracket $\langle[\hat{q}(t), \hat{p} (0)]^2\rangle_\beta$ which can evolve over longer times. Indeed, for some quantum systems the commutator squared may exhibit an exponential growth
\begin{equation}\label{eqQchaos}
\langle[\hat{q}(t), \hat{p} (0)]^2\rangle_\beta\sim e^{\lambda t},
\end{equation}
for time scales $t_{\text{th}}\ll t \ll t_*$, where $t_*\sim \beta \log S$ is a new time scale known as the `Ehrenfest
time' or `scrambling time'. By analogy to the classical case (\ref{eqPBchaos}), such systems are known to be quantum chaotic, with the exponential growth now referring to the spread of the wave function rather than the deviation of nearby trajectories \cite{Larkin1969QuasiclassicalMI,Kitaev:KITP,Maldacena:2016hyu}.

The precise cause of the exponential growth in (\ref{eqQchaos}) can be can be identified as a result of the specific out-of-time-order correlator (OTOC) $\langle q(t)p(0)q(t)p(0)\rangle_\beta$. Recently inspired by the AdS/CFT duality, Maldacena, Shenker and Stanford \cite{Maldacena:2015waa} proved that for systems with a parametrical separation between thermalization and scrambling times, $t_{\text{th}}\ll t_*$, the coefficient $\lambda$ in (\ref{eqQchaos}) should satisfy the bound
\begin{equation}
\label{e20}
\lambda \leq 2 \pi T.
\end{equation}
When the quantum system has a classical gravity dual, one expects $S\sim N^2\gg1$ so the parametrical separation between time scales is satisfied. Quite remarkably, for these systems the inequality (\ref{e20}) is exactly saturated, which is why they are referred to as `maximally chaotic'. In the gravity side, such exponential growth can be understood in terms of the spread of the wavefunction of a particle that falls into a black hole, holographically representing the thermal bath in the dual description. Such growth is also understood microscopically in simple models such as, e.g., the Sachdev-Ye-Kitaev model \cite{Sachdev:1992fk}, which has a sector that is expected to be dual to two-dimensional Jackiw-Teitelboim gravity \cite{Kitaev:2017awl}.

The state of the system, however, keeps on evolving even after scrambling, and it would be desirable to understand and characterize such evolution for quantum chaotic systems. One quantity that has been proposed as a probe of the system's dynamics over these longer time scales is circuit complexity
\begin{equation}
\mathcal{C}=\min g(\hat{U}_{\mt{TR}})\,,
\end{equation}
defined as the minimum number $g$ of elementary gates $\hat{g}_i$ that can be used to construct a quantum circuit $\hat{U}_{\mt{TR}}\sim\Pi_i \hat{g}_i$ that builds a target $\left|\psi_{\mt{T}}\right\rangle$ state from a given reference state $\left|\psi_{\mt{R}}\right\rangle$, i.e., $\left|\psi_{\mt{T}}\right\rangle=\hat{U}_{\mt{TR}}\left|\psi_{\mt{R}}\right\rangle$. Indeed, in the context of quantum computation, this quantity is expected to grow up to a time of order $t_{\text{max}}\sim e^S$, far larger than the scrambling time. At this time, the evolution has taken the system to probe a big portion of the Hilbert space and complexity saturates to its maximum value $\mathcal{C}_{\text{max}}$. Although the aforementioned definition only applies to discrete systems with a finite number of degrees of freedom, research is ongoing to extend the concept of computational complexity to quantum field theories, and there are already several operational definitions of field theory complexity \cite{Jefferson:2017sdb,Chapman:2017rqy,Jiang:2018nzg,Camargo:2018eof,Ali:2018fcz,Ali:2018aon,Chapman:2018hou,Hackl:2018ptj,Khan:2018rzm,Bhattacharyya:2018bbv,Guo:2018kzl,Bhattacharyya:2019kvj,Caceres:2019pgf,Doroudiani:2019llj,Guo:2020dsi}.

Moving on to holography, circuit complexity has been proposed as a diagnose for the growth of the wormhole inside a black hole \cite{Susskind:2014rva,Susskind:2014moa}. There are various proposals for the precise dual of complexity which have been studied extensively: `complexity=volume' (CV) \cite{Stanford:2014jda,Susskind:2014rva}, `complexity=action' (CA) \cite{Brown:2015lvg,Brown:2015bva}, `complexity=spacetime volume' (CV2.0) \cite{Couch:2016exn}, and more recently, their `complexity=anything' generalizations \cite{Belin:2021bga,Belin:2022xmt}. These latter proposals are covariant generalizations of the former quantities that capture some of the basic properties expected in complexity: the linear growth after scrambling and the so called `switchback effect', i.e., the delay in the onset of the linear growth caused by a perturbation (see \cite{Chapman:2021jbh} for a review). Regardless of its definition, holographic complexity is expected to encode precise information about the bulk dynamics and, hence, know about the growth of the wormhole. Crucial work in this direction include \cite{Czech:2017ryf,Caputa:2018kdj,Susskind:2019ddc,Pedraza:2022dqi,Pedraza:2021fgp,Pedraza:2021mkh}, which point towards a novel paradigm in gravitational physics: \emph{gravity emerges from efficient quantum computation}.

\begin{figure}[t!]
\centering
\includegraphics[width=15cm]{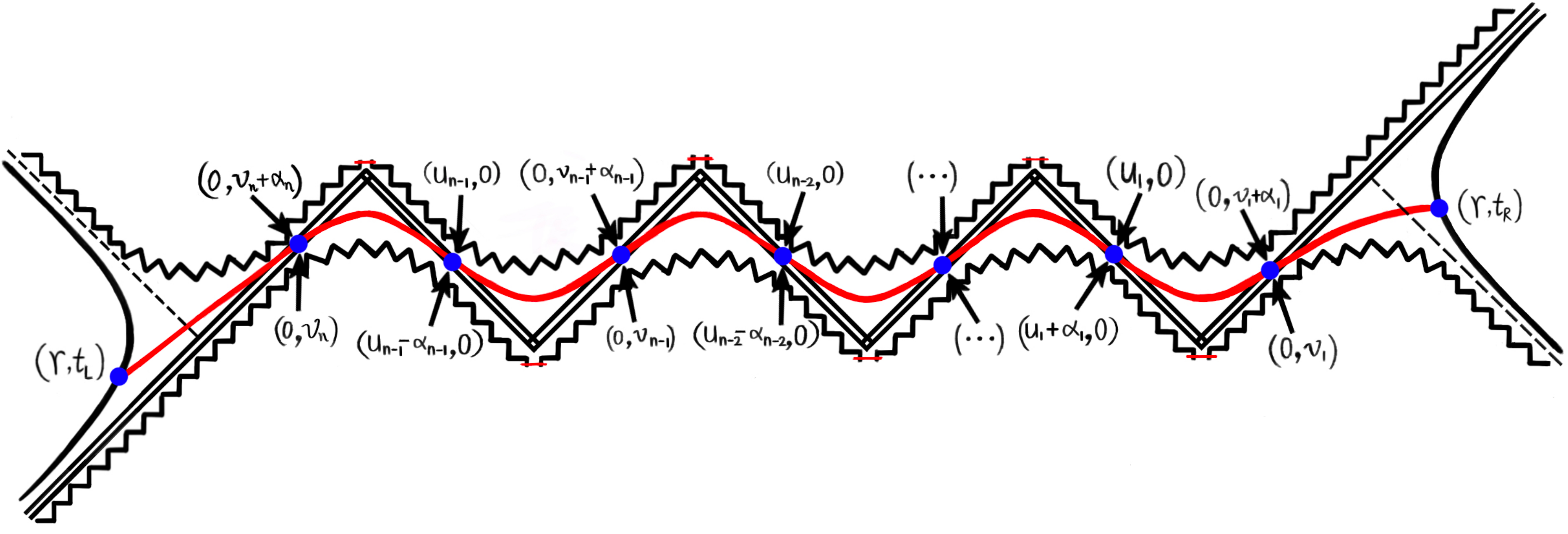}
\caption{Kruskal diagram for an ERB dual to the TFD state that is perturbed at the left boundary by out-of-time-order operators \cite{Stanford:2014jda}. The maximum-volume surface that crosses the ERB is shown by the red curve. There are two sets of $(u,v)$ coordinates for each point of junction with a shock: one in the patch to the left and one in the patch to the right. The intensity of each shock determines the null shifts that link them.}
\label{F10}
\end{figure}

One may wonder if some of the lessons learned about holographic complexity extend to other quantum chaotice systems, not necessarily holographic. As a first thought experiment, one may consider moving beyond large $N$ theories. The intuition here is that the bulk becomes highly quantum and thus, the prescriptions for complexity should receive important $1/N$ corrections. There are some situations, however, in which certain holographic observables may be robust against $1/N$ corrections and hence become `universal'. We would like to engineer a situation in which this may happen for complexity. Inspired by \cite{Stanford:2014jda}, we will consider perturbations of the thermofield double (TFD) state of the form
\begin{equation}
\label{e42}
\left|\Psi\left(t_{\mathrm{L}}, t_{\mathrm{R}}\right)\right\rangle=e^{-i \hat{H}_{\mathrm{L}} t_{\mathrm{L}}-i \hat{H}_{\mathrm{R}} t_{\mathrm{R}}} \hat{W}_\mathrm{L}\left(t_n\right) \ldots \hat{W}_\mathrm{L}\left(t_1\right)\left|{\mathrm{TFD}}\right\rangle,
\end{equation}
where $\hat{W}_\mathrm{L}$ is a `precursor' operator that represents a perturbation to the system and the $t_1,\ldots t_n$ are in an alternating ``zig-zag'' order. The effect of the exponentials in (\ref{e42}) is to implement multiple backward and forward time evolutions of the TFD state, which can be interpreted as a time-fold \cite{Heemskerk:2012mn,Susskind:2013lpa}. As mention above, one property of complexity is the switchback effect. The switchback effect implies that in this setting, the complexity of the system, from here onwards referred to as multifold complexity, should be proportional to \cite{Stanford:2014jda}
\begin{equation}
\label{e43}
\mathcal{C}\propto t_{\text{T}}-2 n t_*,
\end{equation}
where $ t_*$ is the scrambling time of the system and 
\begin{equation}
t_{\text{T}}\equiv\left|t_{\mathrm{R}}+t_1\right|+\left|t_2-t_1\right|+\cdots+\left|t_{\mathrm{L}}-t_n\right|
\end{equation}
is the total folded time. This formula assumes that the time along each switchback (i.e., each term involving an absolute value) is much larger than the scrambling time. For theories with a classical gravity dual, the state \eqref{e42} may be understood in the bulk as a long wormhole supported by alternating left- and right-moving shockwaves. See Fig.~\ref{F10} for an illustration.
Generally calculating a precursor is computationally extremely difficult \cite{Susskind:2013lpa}, especially for systems that are chaotic and with a large number of degrees of freedom. However, in the limit of a large number of precursors, or shock waves, we expect drastic simplifications will happen.
In this limit, the wormhole gets arbitrarily long while the \emph{spacetime} volume supporting it shrinks down. Hence we expect the various holographic proposals for complexity to give qualitatively similar results, being them either codimension-one or codimension-zero as in \cite{Belin:2021bga,Belin:2022xmt}. Furthermore, the spacetime shrinking means that there is little room for quantum fluctuations to become important. Hence, even if we are dealing with finite $N$ systems, we expect the multifold complexity to be largely insensitive to $1/N$ corrections.

To test the above conjecture, namely, the possible universality of the multifold complexity in such a limit, we will study a system with no classical holographic dual, which is nevertheless chaotic. For concreteness, we will examine the inverted harmonic oscillator, a system that has been widely studied in the past in terms of its OTOCs, the Loschmidt echo, and circuit complexity \cite{Roberts:2016hpo,Cotler:2017jue,Magan:2018nmu,Balasubramanian:2019wgd,Yosifov:2019gwt,Ali:2019zcj,Bhattacharyya:2020art,Bhattacharyya:2020iic,Jalabert2001,Hashimoto:2017oit,Hashimoto:2020xfr,Qu:2021ius,Garcia-Mata:2022voo,Bhattacharyya:2021fii,Gao:2022ybw,Chen:2022tbb,Pal:2022rqq,Afrasiar:2022efk,Camblong:2022oet}. The Loschmidt echo examines the inner product of bidirectional evolution. So contrary to our setting, one first evolves the reference state forward with a Hamiltonian $\hat{H}$ and then evolve backward with a slightly different Hamiltonian $\hat{H}+\hat{\delta H}$. Interestingly, we find that both the multifold Loschmidt echo and the multifold complexity have the expected structure as inferred from Eq.~\eqref{e43} and more complex features particular to our model that can be extracted from Eq.~\eqref{rho_L} and \eqref{rho_P}. We find that the latter is the combination of the terms with the shortest scrambling time in the former. Additionally, we prove that multifold complexity is dominated by the longest permutation of the given time combination in an alternating ``zig-zag'' order, obtaining the exact same outcome that Stanford and Susskind predicted in \cite{Stanford:2014jda}.

The rest of the paper is organized as follows. In section 2, we review some background material and preliminaries. In section 3, we discuss the multifold behavior of the inverted harmonic oscillator. Finally, a summary of our findings is presented in section 4.

\section{Background material and preliminaries}
\subsection{Inverted harmonic oscillator}
We begin by considering the Hamiltonian for an inverted harmonic oscillator, \ie
\begin{equation}
\label{e1}
\begin{aligned}
\hat{H}=\frac{\hat{p}^2}{2m}+\hat{V},\quad \hat{V}=-\frac{1}{2} m \omega^2 \hat{q}^2,
\end{aligned}
\end{equation}
where $m$ and $\omega$ stand for the inverted harmonic oscillator's mass and frequency, respectively. An inverted harmonic oscillator's potential $\hat{V}$ is never positive, unlike the ordinary harmonic oscillator. The inverted oscillator lacks a ground state and a full set of square-integrable energy eigenstates because the potential is unbounded. $\hat{V}$ is not a Hermitian operator in the strictest sense of the word because it tends to negative infinity at infinity. However, in most cases, we may consider $\hat{V}$ as local approximation of certain Hermitian operators, such as
\begin{equation}
\hat{V'}=-\frac{1}{2} m \omega^2 \hat{q}^2+\lambda\hat{q}^4.
\end{equation}
It is clear that in the limit $ q \to 0$, $V$ approximates ${V'}$. Alternatively, if the system evolution time is sufficiently short, we may also utilize
$\hat{V}$ instead of $\hat{V'}$ in the Schr\"odinger equation for the state that is localized at the origin. Additionally, as $\hat{V}$ is the quadratic form of $\hat{q}$, its numerous features may be studied analytically. This is another reason why inverted harmonic oscillators are actively researched, with applications ranging from scattering problems, to diffusion in condensed matter systems, black holes and cosmological backgrounds, see e.g. \cite{Barton:1984ey,Seshadri:1989,Maldacena:2005he,Sierra:2008se,Friess:2004tq,Fischler:2014tka,Betzios:2016yaq,Betzios:2020wcv,Berry,Morita:2019bfr,Morita:2018sen,Bzowski:2018aiq,Subramanyan:2020fmx}.

\subsection{Circuit complexity and inner product of Gaussian states}
We quickly go through the definition of circuit complexity first. Readers who need further information might consult Ref.~\cite{Chapman:2018hou}. Through the unitary evolution, we begin with a reference state $\left|\psi_{\mt{R}}\right\rangle$ and prepare a certain target state $\left|\psi_{\mt{T}}\right\rangle$, namely
\begin{equation}
\left|\psi_{\mt{T}}\right\rangle=\hat{U}_{\mt{TR}}\left|\psi_{\mt{R}}\right\rangle.
\end{equation}
We must fix the set of basic gates that can execute the unitary transformation $\hat{U}_{\mt{TR}}$ in order to specify the circuit complexity from the selected reference state to the target state.
The decomposition along a path integral on the group manifold, in turn, relates to a specific method of building the unitary operator $\hat{U}_{\mt{TR}}$, namely
\begin{equation}
\hat{U}_{\mt{TR}}=\overleftarrow{\mathcal{T}} \exp \left[i \int_{0}^{1} \mathrm{~d} t \, Y^I(t)\hat{M_I}\right],
\end{equation}
where $\overleftarrow{\mathcal{T}}$ indicates the path order, $\hat{M_I}$ denotes a set of generators dual to the fundamental gates, and $Y^I(t)$ represents the number of basic gates $\hat{M_I}$ used at ``time'' $t$ along the evolution path, while the choice of the cost functions that parametrize the costs of the various pathways continues to influence the circuit complexity. The simple ones are given by the so-called $F_1$ and $F_2$ cost functions with the circuit complexity as
\begin{equation}
\label{e9}
\begin{split}
\mathcal{C}_1 &=\min _{Y^I(t)}\int_{0}^{1} \mathrm{~d} t \sum_I| Y^I(t)|\,,\\ \mathcal{C}_2&=\min _{Y^I(t)}\int_{0}^{1} \mathrm{~d} t \sqrt{\sum_{I}\left(Y^{I}(t)\right)^{2}}\,.\\
\end{split}
\end{equation}
The least cost of converting a given reference state into a selected target state, or the cost of the optimal quantum circuit, is the circuit complexity, according to the definition.
Therefore, the minimization that appears in the expressions \eqref{e9} refers to the optimization of every route that might go from the reference state to the target state. The shortest geodesic on the group manifold is the only object that makes up the circuit complexity described in \eqref{e9} that is related to the $F_2$ cost function. However, it is difficult to solve a non-trivial group's optimization problem precisely. The calculations for the circuit complexity in the quantum field theory were first explicitly realized in Ref. \cite{Jefferson:2017sdb} for Gaussian states. See Refs. \cite{Chapman:2018hou,Bhattacharyya:2019kvj,Hackl:2018ptj,Khan:2018rzm,Caceres:2019pgf,Bhattacharyya:2018bbv,Guo:2018kzl,Doroudiani:2019llj,Guo:2020dsi} for details on how this progression is being followed.

Because Gaussian states $\ket{\psi}$ are straightforward (see \eg Refs. \cite{Weedbrook:2015bva,wang2007quantum,adesso2012measuring,adesso2014continuous,ferraro2005gaussian}), one may demonstrate that they are entirely specified by the two-point functions described by the so-called covariance matrix,
\begin{equation}
G^{a b} = \bra{\psi}(\hat{\xi}^{a} \hat{\xi}^{b}+\hat{\xi}^{b} \hat{\xi}^{a})\ket{\psi},
\end{equation}
where the vector operator $\hat{\xi}^{a}=\left(\hat{q} g,\frac{\hat{ p}}{g}\right)$. Note that $g$ is a new gate scale, and its dimension is one. We introduce $g$ to offset the dimension of $\hat{q}$ and $\hat{p}$ so that the dimension of $\hat{\xi}^{a}$ is zero. If we take into account the unitary transformation between the Gaussian states,
\begin{equation}
\label{e24}
\begin{aligned}
\ket{\psi^{\prime}}&=\hat{U}\left|\psi\right\rangle
,\quad \hat{U}=e^{-\frac{i}{2} k_{a b}^{} \hat{\xi}^{a} \hat{\xi}^{b}},
\end{aligned}
\end{equation}
As a distinct transformation of their associated covariance matrices, it may be rewritten as follows \cite{Weedbrook:2015bva}
\begin{equation}
\label{Trans}
\begin{aligned}
G^{\prime a b} &=\bra{\psi^{\prime}}(\hat{\xi}^{a} \hat{\xi}^{b}+\hat{\xi}^{b} \hat{\xi}^{a})\ket{ \psi^{\prime}}=M_{\ c}^{a} G^{c d} M_{\ d}^{b},\end{aligned}
\end{equation}
where
\begin{equation}
\label{e25}
\begin{aligned}
M=e^K,\quad    K_{ \ b}^{a} =\Omega^{a c} k_{c b},  \quad \Omega^{a b}=\left(\begin{array}{cc}
0 & 1 \\
-1 & 0
\end{array}\right).
\end{aligned}
\end{equation}
As a result, we should anticipate that the covariance matrix may be used to rewrite the circuit complexity of Gaussian states. Particularly, it was found in Ref. \cite{Chapman:2018hou} that the information about the circuit complexity is encoded in the relative covariance matrix, which is defined by
\begin{equation}
\Delta=G_{\mt{T}}G_{\mt{R}}^{-1}.
\end{equation}
where $G_{\mt{R}}$ and $G_{\mt{T}}$ are the covariance matrix of the reference state and the target state, respectively. The two eigenvalues of the relative covariance matrix $\Delta$ are reciprocal to one another and bigger than zero, as is to be noted \cite{Qu:2021ius}. The following uses the symbol $\rho$ to signify an eigenvalue that is larger than or equal to 1. For Gaussian states, it was explicitly shown in Ref.~\cite{Chapman:2018hou} that the circuit complexity related to the $F_2$ cost function is given by
\begin{equation}
\mathcal{C}_{2}\left(G_{\mathrm{R}}, G_{\mathrm{T}}\right)=\frac{1}{2 \sqrt{2}} \sqrt{\operatorname{Tr}\left[(\log \Delta)^{2}\right]} =\frac{1}{2}\log{\rho}\,,
\end{equation}
The one-mode Gaussian state may be used as an example to show how the geometric technique previously mentioned and the covariance matrix method are equivalent. The unitary transformations between one-mode Gaussian states can be realized by three generators\footnote{In fact, $\hat{q},\hat{ p}$ here means that $G_R$ and $G_T$ are diagonal in $\hat{q},\hat{ p}$, while it does not represent the original $\hat{q},\hat{ p}$ \cite{Chapman:2018hou}.}, {\it viz},
\begin{equation}
\hat{M}_W=\frac{1}{2}(\hat{q}\hat{ p}+\hat{p}\hat{ q}), \quad \hat{M}_V=\frac{\hat{q}^{2}g^2}{\sqrt{2}}, \quad \hat{M}_Z=\frac{\hat{p}^{2}}{\sqrt{2}g^2},
\end{equation}
whose commutation relations precisely match the algebra $\mathfrak{s p}(2, \mathbb{R})$. It was shown in Ref.~\cite{Chapman:2018hou} that the shortest geodesic line for $\mathcal{C}_1$ and $\mathcal{C}_2$ is the straightline parametrized by
\begin{equation}
Y^{W}=-\frac{1}{2}\log\rho, \quad Y^{V}=0, \quad Y^{Z}=0.
\end{equation}
When the optimum path's solution is included in the definitions of circuit complexity, \ie Eq.~\eqref{e9}, we easily obtain
\begin{equation}
\label{e28}
\begin{aligned}
\mathcal{C}_1&=
\int_{0}^{1} \mathrm{~d} t \sum_I| Y^I(t)|=\frac{1}{2}\log{\rho},   \\
\mathcal{C}_2 &=\int_{0}^{1} \mathrm{~d} t \sqrt{\sum_{I}\left(Y^{I}(t)\right)^{2}}=\frac{1}{2}\log{\rho},
\end{aligned}
\end{equation}
which is the same as that derived from the covariance matrix.

On the other hand, we are equally curious about the inner product between the reference state and the target state. The inner product for the Gaussian states also depends on the relative covariance matrix, \ie
\begin{equation}
\label{e8}
\mathcal{I}=\left|\left\langle G_{\mt{R}} \mid G_{\mt{T}}\right\rangle\right|^{2}=\operatorname{det} \frac{\sqrt{2} \Delta^{1 / 4}}{\sqrt{\mathbb{1}+\Delta}}=\frac{2 \sqrt{\rho}}{1+\rho}.
\end{equation}
From this vantage point, we can observe that the circuit complexity between states and the inner product of the Gaussian states convey the same information. When $\rho$ is much larger than one, we have
\begin{equation}
-\log\mathcal{I}=-\log\frac{2 \sqrt{\rho}}{1+\rho}\approx\frac{1}{2}\log\rho-\log 2.
\end{equation}
Since both the complexity and the inner product are determined by $\log\rho$, we only discuss $\log\rho$ in the following.

%%%%%%%%%%%%%%%%%%%%%%%%%%%%%
\section{Multifold complexity for an inverted harmonic oscillator}
In this section, we investigate the chaos in inverted harmonic oscillators by using the circuit complexity and the inner product. Firstly, we choose the reference state as
\begin{equation}
\psi_{\mt{R}}\left(q\right)=\left(\frac{  m\omega}{\pi}\right)^{1/4} 
  e^{- \frac{1}{2} m\omega q^{2}},
\end{equation}
It is obvious that the reference state is nothing but the ground state of the harmonic oscillator with mass $m$ and frequency $\omega$. Correspondingly, its two-point function reads as
\begin{equation}
G_{\mt{R}}=\left(\begin{array}{cc}
\frac{g^2}{m\omega} &0\\
0 & \frac{m\omega}{g^2}  \\
\end{array}\right).
\end{equation}

Inspired by Loschmidt echo \cite{Jalabert2001}, we consider the Loschmidt state $|\psi_{\mt{L}}\rangle$ by multiple backward and forward time evolutions which is not typically in time-order from the reference state, \ie
\begin{equation}
\label{e30}
|\psi_{\mt{L}}\rangle=e^{i \hat{H}'  t_N}e^{-i \hat{H}  t_N}\cdots e^{i \hat{H}'  t_3}e^{-i \hat{H}  t_3}
e^{i \hat{H}'  t_2}e^{-i \hat{H}  t_2} e^{i \hat{H}'  t_1}e^{-i \hat{H}  t_1}\left|\psi_{\mt{R}}\right\rangle\,,
\end{equation}
where $\hat{H}$ is the Hamiltonian of the inverted Harmonic Oscillator, and $\hat{H}'=\hat{H}+\delta\hat{H}$ is defined by
\begin{equation}\label{eq:perturbedH02}
\hat{H}'=\frac{\hat{p}^2}{2m}-\frac{1}{2} m (\omega+\delta\omega)^2 \hat{q}^2\,\quad  \text{with}\qquad  \delta\omega\ll{\omega}\,.
\end{equation}
Obviously, without the perturbation $\delta \hat{H}$, the Loschmidt state trivially evolves back to the reference state after the multiple backward and forward time evolutions. For a chaotic system, if $\delta \hat{H}$ is an operator that does not commute with the Hamiltonian $\hat{H}$, The inner product of $|\psi_{\mt{L}}\rangle$ and $\left|\psi_{\mt{R}}\right\rangle$ will decrease over time \cite{Jalabert2001}. This is the embodiment of the butterfly effect, that is, the accumulation of small perturbations over time leads to huge changes.

On the other hand, inspired by shock wave geometries \cite{Stanford:2014jda}, we also consider a precursor state $\left|\psi_{\mt{P}}\right\rangle$ which is obtained by applying the precursor operator multiple times to the reference state, \ie
\begin{equation}
|\psi_{\mt{P}}\rangle=e^{-i \hat{H}  t_f}\hat{W}(t_N)\cdots \hat{W}(t_3)\hat{W}(t_2)\hat{W}(t_1) e^{i \hat{H}  t_s}\left|\psi_{\mt{R}}\right\rangle\,,
\end{equation}
where $\hat{W}(t)=\hat{U}^{\dagger}(t) \hat{W} \hat{U}(t)$ is the precursor operator of the form
\begin{equation}
\label{e31}
\hat{W}(t)=e^{i \hat{H}  t} e^{-\frac{i}{2} m\delta\omega  \hat{q}^2} e^{-i \hat{H}  t}\,\quad  \text{with}\qquad  \delta\omega\ll{\omega}\,,
\end{equation}
where $\delta\omega\ll \omega$ is an analogy for local requirements in a multi-degree-of-freedom system, \ie $\hat{W}$ needs to be very close to the unit operator for the whole system, $t_s$ and $t_f$ are the results of the evolution of the TFD state reduced to one side, \ie $t_s=-t_\mathrm{R},t_f=t_\mathrm{L}$. Note that if $\hat{W}$ is exactly the unit operator, the $\hat{U}^{\dagger}(t)$ will cancel with the $\hat{U}(t)$, then $\hat{W}(t)$ is also the unit operator. Conversely, if $\hat{W}$ does not commute with the Hamiltonian $\hat{H}$, the size of $\hat{W}(t)$ will grow with time although $\hat{W}$ is very small \cite{Susskind:2014rva}. Like the state $\left|\psi_{\mt{L}}\right\rangle$, the state $\left|\psi_{\mt{P}}\right\rangle$ also reflects the butterfly effect at the quantum level\footnote{ The main difference between the two states is whether the perturbation increases over time.}.

\subsection{Main results}
In order to calculate the complexity and inner product, we first need to calculate the eigenvalues of the relative covariance matrix $\Delta$. For a single inverted harmonic oscillator if we assume
\begin{equation}
\begin{aligned}
\Delta=\left(\begin{array}{cc}
\Delta_{11} & \Delta_{12} \\
\Delta_{21} & \Delta_{22}
\end{array}\right),
\end{aligned}
\end{equation}
then we have
\begin{equation}
\begin{aligned}
\rho=\frac{1}{2}(\Delta_{11}+\Delta_{22}+\sqrt{\Delta_{11}^2+4\Delta_{12}\Delta_{21}-2\Delta_{11}\Delta_{22}+\Delta_{22}^2}),\\
\frac{1}{\rho}=\frac{1}{2}(\Delta_{11}+\Delta_{22}-\sqrt{\Delta_{11}^2+4\Delta_{12}\Delta_{21}-2\Delta_{11}\Delta_{22}+\Delta_{22}^2}).
\end{aligned}
\end{equation}
Since we care about the exponentially growing behavior of the system, i.e. exponentially growing $\rho$ and exponentially decreasing $\frac{1}{\rho}$, in the leading order we have
\begin{equation}
\label{eq1}
\begin{aligned}
\rho&=\frac{1}{2}(\Delta_{11}+\Delta_{22}+\sqrt{\Delta_{11}^2+4\Delta_{12}\Delta_{21}-2\Delta_{11}\Delta_{22}+\Delta_{22}^2})\\
& \approx \frac{1}{2}(\Delta_{11}+\Delta_{22}+\Delta_{11}+\Delta_{22})\\
& = \Delta_{11}+\Delta_{22}.
\end{aligned}
\end{equation}
We find that the eigenvalues depend linearly on the matrix components in the leading order. Furthermore, combining Eqs.~\eqref{Trans}, \eqref{e30} and \eqref{e31}, we can find that the fastest rate of exponential growth is $4\omega$ for single $t$ since each $M$ contributes one $\omega t$ and we have four $M$s in the final formula. Therefore, we can ignore other exponentially growing terms inside the logarithmic function and only keep the dominant term $\exp|4\omega t|$ since such terms will dominate in the late time.

\begin{figure}[t!]
\centering
\includegraphics[width=15cm]{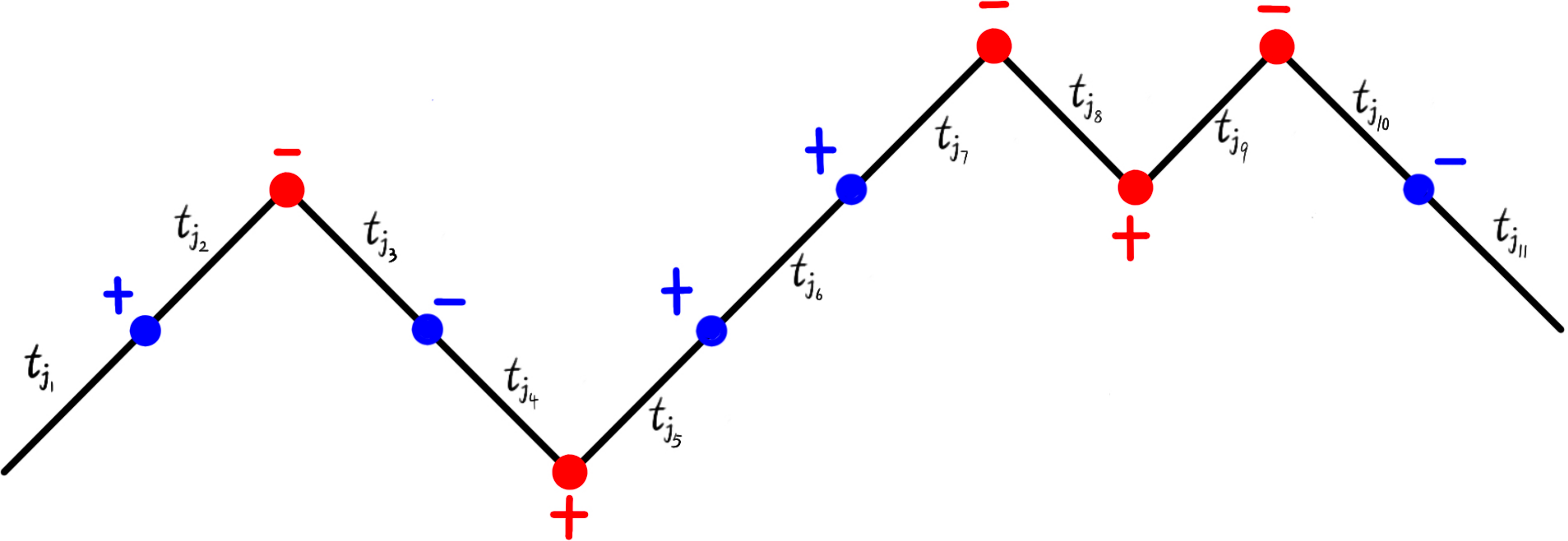}
\caption{Schematic diagram of $\kappa$. Line segments represent different times, and they are connected by point (plus or minus). The blue point means the sign has not changed, while the red point means the sign has changed. There are five red points in the picture, therefore we have $\kappa=5$.}
\label{FF2}
\end{figure}

At the end of this subsection, we give the general expression for $N\geq 1$ and discuss it in the following sections. In order to do this, we need to define a new symbol $\kappa(\oplus_1,\oplus_2,\cdots,\oplus_{n-1})$, where $\oplus_i$ can be plus or minus. This symbol describes the number of the sign changed in the arrangement, \ie
\begin{equation}
\kappa(+)=0,\kappa(-)=\kappa(+,-)=\kappa(-,-)=1,\kappa(-,+)=\kappa(-,-,+)=2.
\end{equation}
For an illustration, see Fig.~\ref{FF2}.
Using this symbol, the eigenvalue of the relative covariance matrix for the Loschmidt state is given by
%\begin{equation}
%\label{rho_L}
%\begin{aligned}
%\rho_\mt{L} &\approx \rho_\mt{L}^0=e^ {|2 t_f \omega|}\\
%&+\sum_{n=1}^{N}\sum_{j_1<j_2<\cdots<j_n}\sum_{\oplus_1,\cdots,\oplus_{n-1}}\left(\frac{\delta \omega^2}{4\omega^{2}}\right)^{2(n-1)-\kappa(\oplus_1,\cdots,\oplus_{n-1})}\frac{\delta \omega^2 \cosh[4 \big{(}t_{j_1}\oplus_1 t_{j_2}\oplus_2  \cdots\oplus_{n-1} t_{j_n}-(\oplus_{n-1}\frac{t_f}{2})\big{)} \omega]}{2 \omega^2}\\
%&+\sum_{n=1}^{N}\sum_{j_1<j_2<\cdots<j_n}\sum_{\oplus_1,\cdots,\oplus_{n-1}}\left(\frac{\delta \omega^2}{4\omega^{2}}\right)^{2n-1-\kappa(\oplus_1,\cdots,\oplus_{n-1})}\frac{\delta \omega^2 \cosh[4 (t_{j_1}\oplus_1 t_{j_2}\oplus_2 \cdots\oplus_{n-1} t_{j_n}\oplus_{n-1}\frac{t_f}{2}) \omega]}{2 \omega^2},
%\end{aligned}
%\end{equation}
\begin{equation}
\label{rho_L}
\begin{aligned}
\rho_\mt{L} \approx \rho_\mt{L}^0 =1+ \sum_{n=1}^{N} ~\sum_{j_1<j_2<\cdots<j_n}~ \sum_{\oplus_1,\cdots,\oplus_{n-1}}
    \alpha^{2n-1-\kappa(\oplus_1,\cdots,\oplus_{n-1})}
    e^{\left|4 (t_{j_1}\oplus_1 t_{j_2}\oplus_2  \cdots\oplus_{n-1} t_{j_n}
                 ) \omega\right|}
               ,
\end{aligned}
\end{equation}
where $\alpha=\frac{\delta \omega^2}{4\omega^{2}}$ and $j_k\in\{1,2,\cdots,N\}$. On the other hand, the eigenvalue of the relative covariance matrix for the precursor state is
given by
\begin{equation}
\label{rho_P}
\begin{aligned}
\rho_\mt{P} &\approx \rho_\mt{P}^0 = e^ {|2 (t_s-t_f) \omega|}
   + \sum_{n=1}^{N} \sum_{j_1<j_2<\cdots<j_n}
   \alpha^{n}
     e^{\left|4 \big( \frac{t_s}{2}+ \sum_{k=1}^{n} (-1)^{k}t_{j_k}
        +(-1)^{n+1}\frac{t_f}{2}\big) \omega
            \right|}.
\end{aligned}
\end{equation}
The first term is the embodiment of one-way evolution without the butterfly effect. The second term is nothing but the arrangement that maximizes $\kappa$ in Eq.~\eqref{rho_L}, \ie the sign flips every time. Note that on the one hand, the coefficients will become smaller and smaller as more and more different times are included in Eq.~\eqref{rho_L} and \eqref{rho_P}, and on the other hand, the coefficients will increase as the number of the sign is changed. The negative logarithm of the coefficients represents the scrambling time for each term, which is when the complexity and inner product start to change \cite{Qu:2021ius}. This shows that the effects of multiple disturbances need a longer scrambling time to be reflected, and the opposite evolution time can reduce the scrambling time.

If the coefficient of each term is recorded as $\alpha^\sigma$, we can intuitively see the physical meaning of $\sigma$ through the expression of the Loschmidt state
\begin{equation}
|\psi_{\mt{L}}\rangle=e^{-i \hat{H}  t_f}e^{i \hat{H}'  t_N}e^{-i \hat{H}  t_N}\cdots e^{i \hat{H}'  t_3}e^{-i \hat{H}  t_3}
e^{i \hat{H}'  t_2}e^{-i \hat{H}  t_2} e^{i \hat{H}'  t_1}e^{-i \hat{H}  t_1} \left|\psi_{\mt{R}}\right\rangle.
\end{equation}
Note that since we are examining bidirectional evolution, that is, the superposition of forward evolution and backward evolution. For each $t_j$, forward evolution and reverse evolution must be in competition with each other, and scrambling time is the embodiment of competition. But for the adjacent two items, if the time is opposite, the relationship of competition will change to the relationship of cooperation, \ie
\begin{equation}
e^{i \hat{H}'  t_{j+1}}e^{-i \hat{H}  t_{j+1}} e^{i \hat{H}'  t_j}e^{-i \hat{H}  t_j}=e^{-i \hat{H}'  t_j}e^{i \hat{H}  t_j} e^{i \hat{H}'  t_j}e^{-i \hat{H}  t_j}, \quad \text{if} \quad t_{j+1}=-t_j.
\end{equation}
So the above formula contains two competitions on the side and one cooperation in the middle, and the corresponding term is $\alpha^2 \exp|t_j-t_{j+1}|$. Conversely, if $t_{j+1}$ and $t_j$ have the same sign, the corresponding term is $\alpha^3 \exp|t_j+t_{j+1}|$. Therefore $\sigma$ represents the number of competition times, and the corresponding scrambling time is $-\sigma\frac{\log\alpha}{4\omega}$. This makes sense, because the more competition, the longer the corresponding scrambling time. This is also the reason why the symbol $\kappa$ appears. Note that the single scrambling time is 
\begin{equation}
t_*=-\frac{\log\alpha}{4\omega},
\end{equation}
so the total scrambling time is $\sigma t_*$. This is very similar to Eq.~\eqref{e43}, but $n$ in Eq.~\eqref{e43} represents the number of switchbacks, and here we only care whether the adjacent evolutions are in the same direction. On the other hand, adjacent evolutions are generated by the same Hamiltonian for the precursor state, that is, they cancel out completely if the signs are the same, \ie
\begin{equation}
\label{e315}
\begin{aligned}
&e^{i \hat{H}  t_{j+1}} e^{-\frac{i}{2} m\delta\omega  \hat{q}^2} e^{-i \hat{H}  t_{j+1}}e^{i \hat{H}  t_j} e^{-\frac{i}{2} m\delta\omega  \hat{q}^2} e^{-i \hat{H}  t_j}\\
=&e^{i \hat{H}  t_{j+1}} e^{-\frac{i}{2} m\delta\omega  \hat{q}^2} e^{i \hat{H}  (t_j-t_{j+1})} e^{-\frac{i}{2} m\delta\omega  \hat{q}^2} e^{-i \hat{H}  t_j}\\
=&e^{i \hat{H}  t_{j}} e^{-{i} m\delta\omega  \hat{q}^2} e^{-i \hat{H}  t_j},
\end{aligned}
\end{equation}
in the last step, we make $t_{j+1}=t_j$. It can be seen that terms with the same sign degenerate, so we only get terms with the opposite sign in Eq.~\eqref{rho_P}.

In fact, since the complexity and inner product have the form $$\log\big(e^{t_1-t_*}+\cdots+e^{t_1+t_2+\cdots+t_n-(2n-1)t_*}\big),$$ 
the largest exponential term at each instant will completely dominate the behavior of the logarithm. Therefore, we only need to calculate the contribution of the term that maximizes the absolute value of the permutation of the given time combination minus the corresponding scrambling time, and other terms can be ignored at this time combination. For example, for the Loschmidt state, if we consider a double perturbation, we need to compare $|t_1|-t_*,|t_2|-t_*,|t_1-t_2|-2t_*,|t_1+t_2|-3t_*$. In most cases, there will be only one dominance, but for the intersection of the above lines, \ie points with multiple maxima. There will be multiple simultaneous dominance at these points. Still, if we deviate from these points we find that things quickly revert to single-term dominance due to the fast-growing nature of the exponential function.

For the precursor state, the dominant term is the longest permutation of the given time combination in an alternating ``zig-zag'' order. This is because this term represents the longest contour, and the term containing the through-going point represents the difference between the two segments separated by the through-going point.
\begin{figure}[t!]
\centering
\includegraphics[width=2cm]{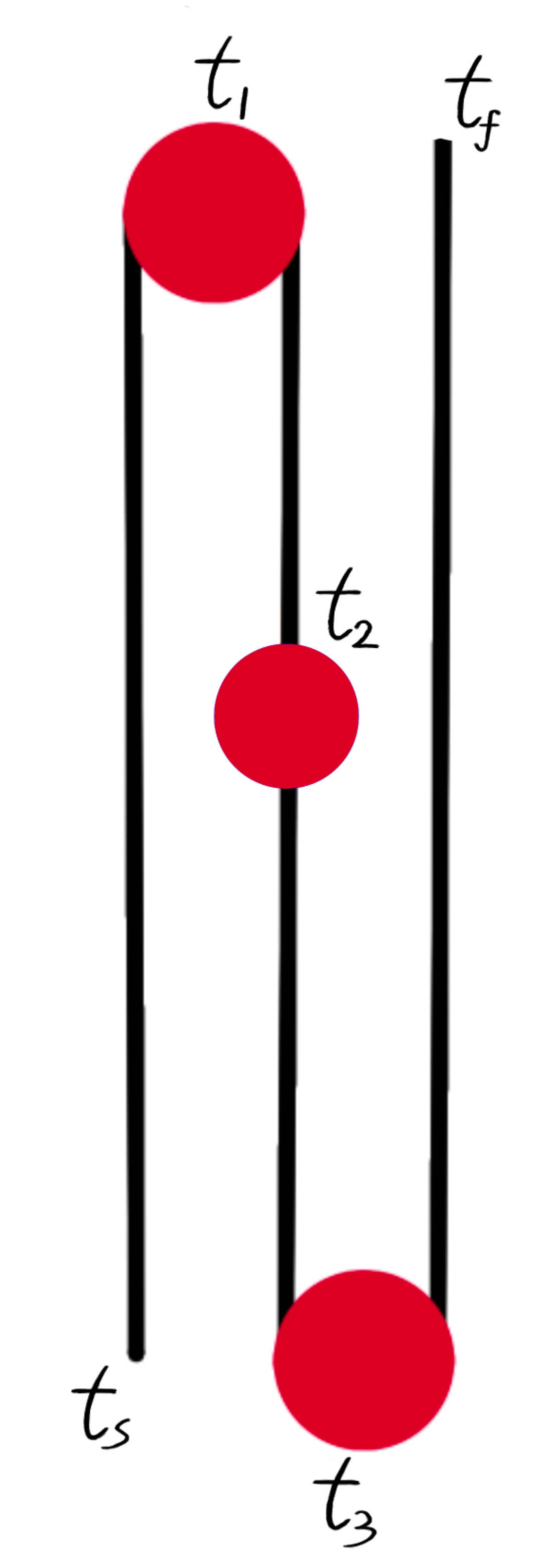}
\caption{ Time-fold with three insertions.
The insertions at $t_1,t_3$ occur at the switchback points. The insertion at $t_2$ occurs at the through-going point.}
\label{FFF1}
\end{figure}
For example, the longest permutation in Fig.~\ref{FFF1} is $t_s,t_1,t_3,t_f$, corresponding to
\begin{equation}
    \frac{t_s}{2}-t_1+t_3-\frac{t_f}{2}=\frac{1}{2}\Big[(t_s-t_1)+(t_3-t_1)+(t_3-t_f)\Big].
\end{equation}
It is nothing but the total length of the polylines in Fig.~\ref{FFF1}. However, if we consider the term with the through-going point $t_s,t_1,t_2,t_3,t_f$,  corresponding to
\begin{equation}
    \frac{t_s}{2}-t_1+t_2-t_3+\frac{t_f}{2}
    =\frac{1}{2}\Big[\big((t_s-t_1)+(t_2-t_1)\big)-\big((t_3-t_2)+(t_3-t_f)\big)\Big]
\end{equation}
It is nothing but the difference between the two segments separated by $t_2$. Obviously, the total length of a line segment is always greater than the difference between the parts. Conversely, if the selected one is not the longest permutation in an alternating ``zig-zag'' order, it is always possible to add a new switchback point to make the total length longer, since a straight line between two points is always shorter than a polyline. Here we ignore the effect of scrambling time because we assume that each line length is greater than scrambling time \cite{Stanford:2014jda}. Therefore, the dominant term is always the longest permutation of the given time combination in an alternating ``zig-zag'' order, \ie $\frac{1}{2}(|t_s-t_1|+|t_1-t_{2}|+\cdots+|t_{n-1}-t_{n}|+|t_{n}-t_{f}|)=\frac{1}{2}t_\text{T}$ with $t_s=-t_\mathrm{R},t_f=t_\mathrm{L}$.
Finally, the corresponding complexity is
\begin{equation}
\mathcal{C}= \omega (t_{\text{T}}-2 n t_*),
\end{equation}
this is exactly the result of the Douglas Stanford and Leonard Susskind conjectures \cite{Stanford:2014jda}. This also proves that Eq.~\eqref{e43} does hold for more general systems, not just for the system with classical holographic dual.

It is worth mentioning that the analytical results we have obtained fit well for a late time and are biased for some early time because some ignored terms come into play at turning points. This can be reflected in the following pictures.

\subsection{Single perturbation}
In this subsection, we discuss in detail the case of the Loschmidt state and precursor state with only one perturbation, namely $N=1$. As discussed in Eq.~\eqref{e24}, we can represent the unitary operators $e^{-i \hat{H}  t_1}$ and $e^{i \hat{H}'  t_1}$ by the matrices %$k(t_1),k^{\prime}(t_1)$
\begin{equation}
k(t)=\left(\begin{array}{cc}
-\frac{\omega^2 m t}{g^2} & 0 \\
0 & \frac{g^2 t}{m}
\end{array}\right) \quad \text{and }\quad  k^{\prime}(t)=\left(\begin{array}{cc}
\frac{(\omega+\delta \omega)^2 m t}{g^2} & 0 \\
0 & -\frac{g^2 t}{m}
\end{array}\right),
\end{equation}
respectively. We can obtain the covariance matrix of the Loschmidt state by the transformation of the covariance matrix shown in Eq.~\eqref{Trans}, \ie
\begin{equation}
G_{\mt{L}}=M'(t_1)M(t_1)G_{\mt{R}}M(t_1)^TM'(t_1)^T,
\end{equation}
where
\begin{equation}
\begin{aligned}
M(t)&=\left(\begin{array}{cc}
\frac{1}{2} e^{-\omega t}\left(e^{2 \omega t}+1\right) & \frac{e^{-\omega t}\left(e^{2 \omega t}-1\right) g^{2}}{2 m \omega} \\
\frac{e^{-\omega t}\left(e^{2 \omega t}-1\right) m \omega}{2 g^{2}} & \frac{1}{2} e^{-\omega t}\left(e^{2 \omega t}+1\right)
\end{array}\right),
\\
M'(t)&=\left(\begin{array}{cc}
\frac{1}{2} e^{-t(\omega+\delta \omega)}\left(e^{2(\omega+\delta \omega)t}+1\right) & -\frac{e^{-t(\omega+\delta \omega)}\left(e^{2(\omega+\delta \omega)t}-1\right) g^{2}}{2 m(\omega+\delta \omega)} \\
-\frac{e^{-t(\omega+\delta \omega)}\left(e^{2(\omega+\delta \omega)t}-1\right) m(\omega+\delta \omega)}{2 g^{2}} & \frac{1}{2} e^{-t(\omega+\delta \omega)}\left(e^{2(\omega+\delta \omega)t}+1\right)
\end{array}\right).
\end{aligned}
\end{equation}
The eigenvalue $\rho_\mt{L}$ of the relative covariance matrix $\Delta_{\mt{L}}=G_LG_R^{-1}$ at leading order is given by \cite{Qu:2021ius}
\begin{equation}
\label{e37}
\begin{aligned}
\rho_\mt{L} &\approx \rho_\mt{L}^0
 =
  %+\frac{\delta \omega^2 }{2 \omega^2} \cosh\Big[4 (\mathrm{t_1}-\frac{t_f}{2}) \omega \Big]
  %+\frac{\delta \omega^4 }{8 \omega^4} \cosh\Big[4(\mathrm{t_1}+\frac{t_f}{2}) \omega \Big].
  1+ \alpha\exp|4\omega t_{1}|
\end{aligned}
\end{equation}
The above formula is exactly the special case of Eq.~\eqref{rho_L} with $N=1$.
\begin{figure}[t!]
\centering
\includegraphics[width=3in]{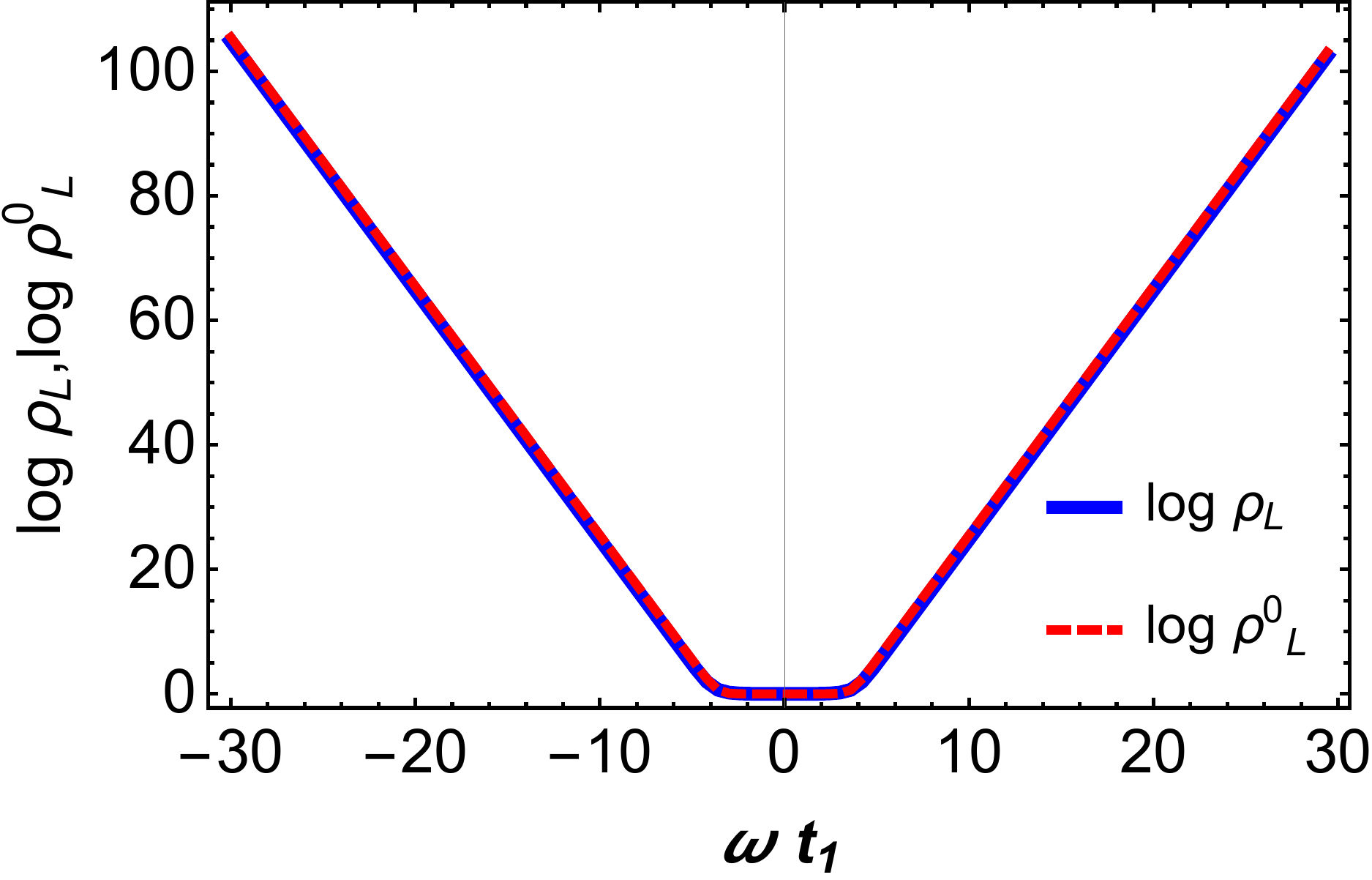}
\includegraphics[width=3in]{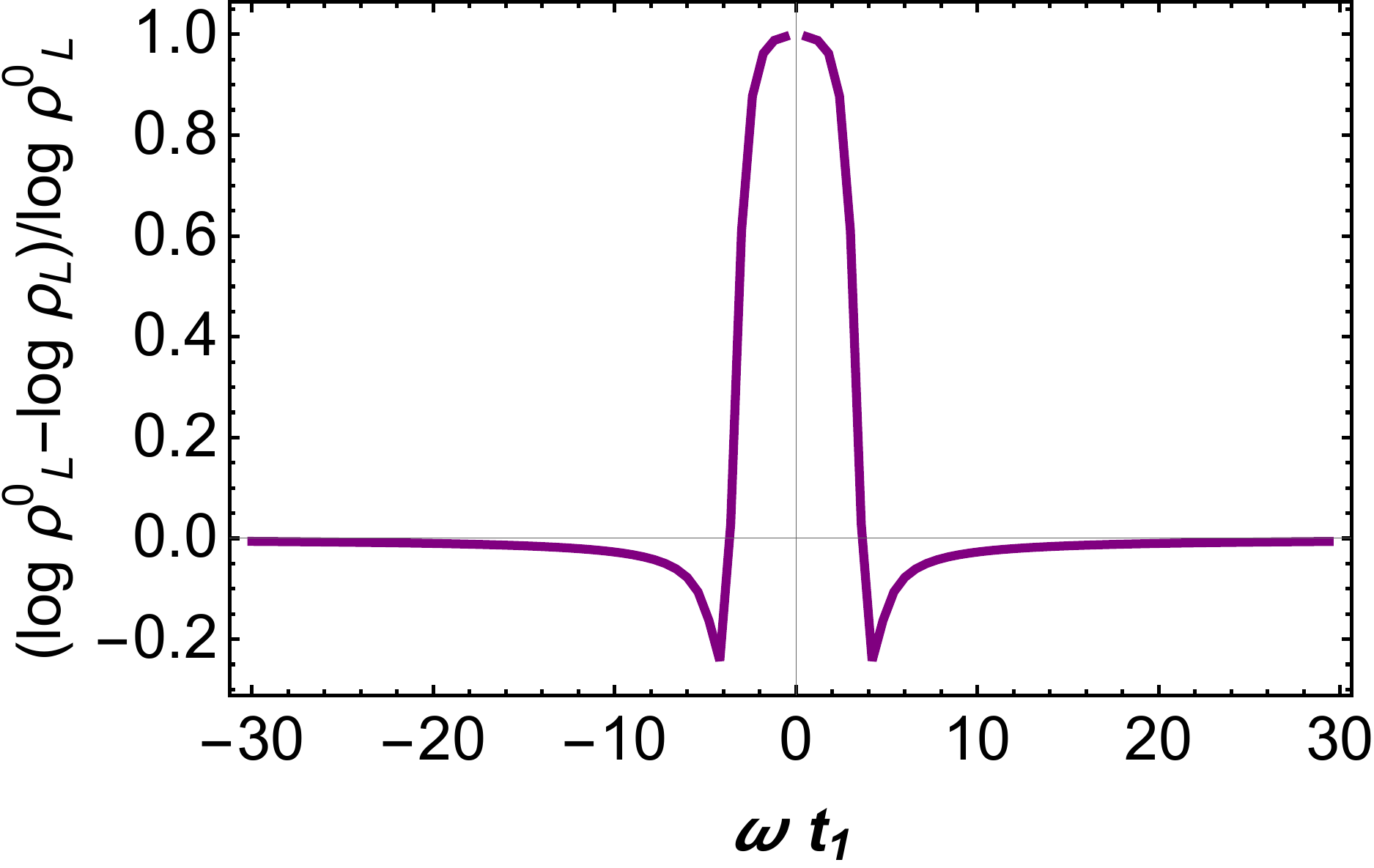}
\caption{$\log\rho_\mt{L}$ and $\log\rho_\mt{L}^0$ vs time for the figure on the left. Relative error $(\log\rho_\mt{L}^0-\log\rho_\mt{L})/\log\rho_\mt{L}^0$ vs time for the figure on the right. The parameters is set as $\frac{\delta\omega}{\omega}=10^{-3}$.}
\label{F3}
\end{figure}
Note that Eq.~\eqref{e37} is exactly the result of the leading order in Ref.~\cite{Qu:2021ius}. Comparison between the precise
value of $\log\rho_\mt{L}$ and $\log\rho_\mt{L}^0$ is shown in Fig.~\ref{F3}.

Before discussing the precursor state of the inverted harmonic oscillator, let us first examine the precursor state of the harmonic oscillator, \ie we change the Hamiltonian to be
\begin{equation}
\begin{aligned}
\hat{H}_h=\frac{\hat{p}^2}{2m}+\frac{1}{2} m \omega^2 \hat{q}^2.
\end{aligned}
\end{equation}
As introduced around Eq.~\eqref{e24}, we  can parametrize the Gaussian unitary operators $e^{-i \hat{H}_h  t_1}$ and $e^{-\frac{i}{2} m\delta\omega  \hat{q}^2}$ by the symmetric matrices
\begin{equation}
k_h(t)=\left(\begin{array}{cc}
\frac{\omega^2 m t}{g^2} & 0 \\
0 & \frac{g^2 t}{m}
\end{array}\right)\quad \text{and} \quad w=\left(\begin{array}{cc}
\frac{m\delta\omega}{g^2}  & 0 \\
0 & 0
\end{array}\right),
\end{equation}
respectively. Since $G_{\mt{R}}$ is an eigenstate of $\hat{H}_h$, we change the reference state to be $G_{\mt{R}}^h$, \ie
\begin{equation}
G_{\mt{R}}^h=\left(\begin{array}{cc}
\frac{g^2}{2m\omega} &0\\
0 & \frac{2m\omega}{g^2}  \\
\end{array}\right).
\end{equation}
It is nothing but the ground state of a harmonic oscillator with frequency $2\omega$ and mass $m$. We can obtain the covariance matrix of the precursor state by the transformation of the covariance matrix shown in Eq.~\eqref{Trans}, \ie
\begin{equation}
G_{\mt{P}}^h=M_h(-t_1)WM_h(t_1)G_{\mt{R}}^hM_h(t_1)^TW^TM_h(-t_1)^T,
\end{equation}
where
\begin{equation}
M_h(t)=\left(\begin{array}{cc}
\frac{1}{2} e^{-i \omega t}\left(e^{2 i t \omega}+1\right) & -\frac{i e^{-i \omega t}\left(e^{2 i \omega t}-1\right) g^2}{2 m \omega} \\
\frac{i e^{-i \omega t}\left(e^{2 i \omega t}-1\right) m \omega}{2 g^2} & \frac{1}{2} e^{-i \omega t}\left(e^{2 i \omega t}+1\right)
\end{array}\right),\quad W=\left(\begin{array}{cc}
1  & 0 \\
-\frac{m\delta\omega}{g^2} & 1
\end{array}\right).
\end{equation}
It is easy to find the eigenvalue $\rho_\mt{P}^h$ of the relative covariance matrix $\Delta_{\mt{P}}^h=G_P^hG_R^{-1}$, \ie
\begin{equation}
\begin{aligned}
\rho_\mt{P}^h&=1
   +\frac{\delta\omega^2}{32\sqrt{2} \omega^2} \Big [
    \sqrt{2}\left({25}-30 \cos(2  \omega t_ 1 )
         +9 \cos^2(2  \omega t_ 1 ) \right)
\\
&+   %\frac{\delta\omega^2}{32 \sqrt{2} \omega^2} 
  \big(5-3 \cos (2  \omega t_ 1 )\big)
  \sqrt{59-60 \cos(2  \omega t_ 1 )+9  \cos(4  \omega t_ 1 ) +128 \omega^2/ \delta \omega^2} 
  \Big ].
\end{aligned}
\end{equation}
Since the perturbation tends to zero, the leading order of $\rho_P$ is always one, that is to say, the influence of the initial perturbation on the system can be ignored, which is consistent with the intuition that the harmonic oscillator is not a chaotic system \cite{Qu:2021ius}.

Thanks to the previous preparations, we can easily get the covariance matrix of the precursor state for the inverted harmonic oscillator
\begin{equation}
G_{\mt{P}}=M(t_f)M(-t_1)WM(t_1)M(t_s)G_{\mt{R}}M(t_s)^TM(t_1)^TW^TM(-t_1)^TM(t_f)^T.
\end{equation}
The eigenvalue $\rho_\mt{P}$ of the relative covariance matrix $\Delta_{\mt{P}}=G_PG_R^{-1}$ at leading order is given by
\begin{equation}
\label{e39}
\begin{aligned}
\rho_\mt{P} &\approx \rho_\mt{P}^0
  =e^ {|2 (t_f-t_s) \omega|}
   +\alpha e^{\big|4(\frac{t_s}{2}-{t_1}+\frac{t_f}{2})\omega\big|}.
  \end{aligned}
\end{equation}
The above formula is exactly Eq.~\eqref{rho_P} with $N=1$.
\begin{figure}[t!]
\centering
\includegraphics[width=3in]{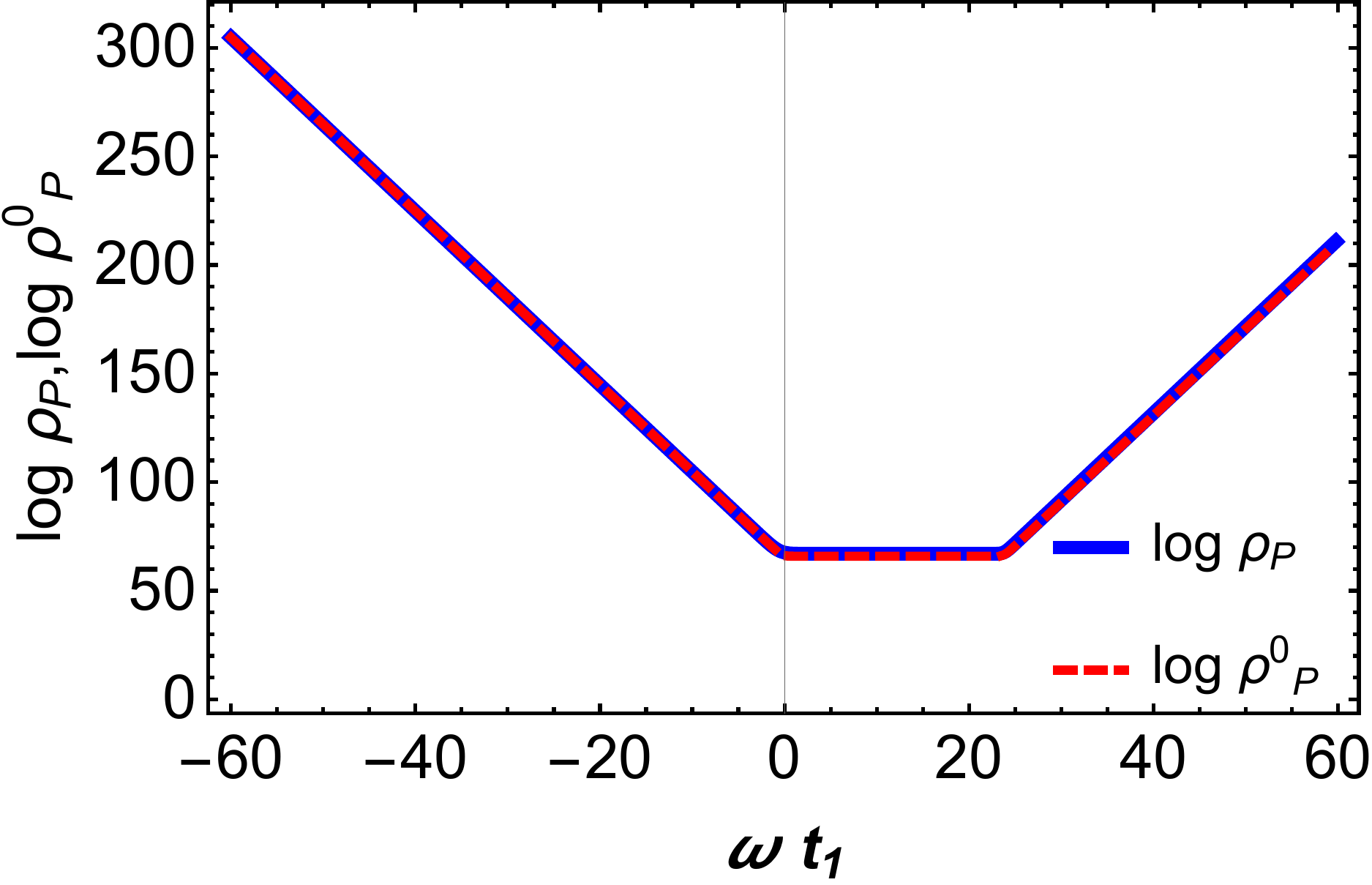}
\includegraphics[width=3in]{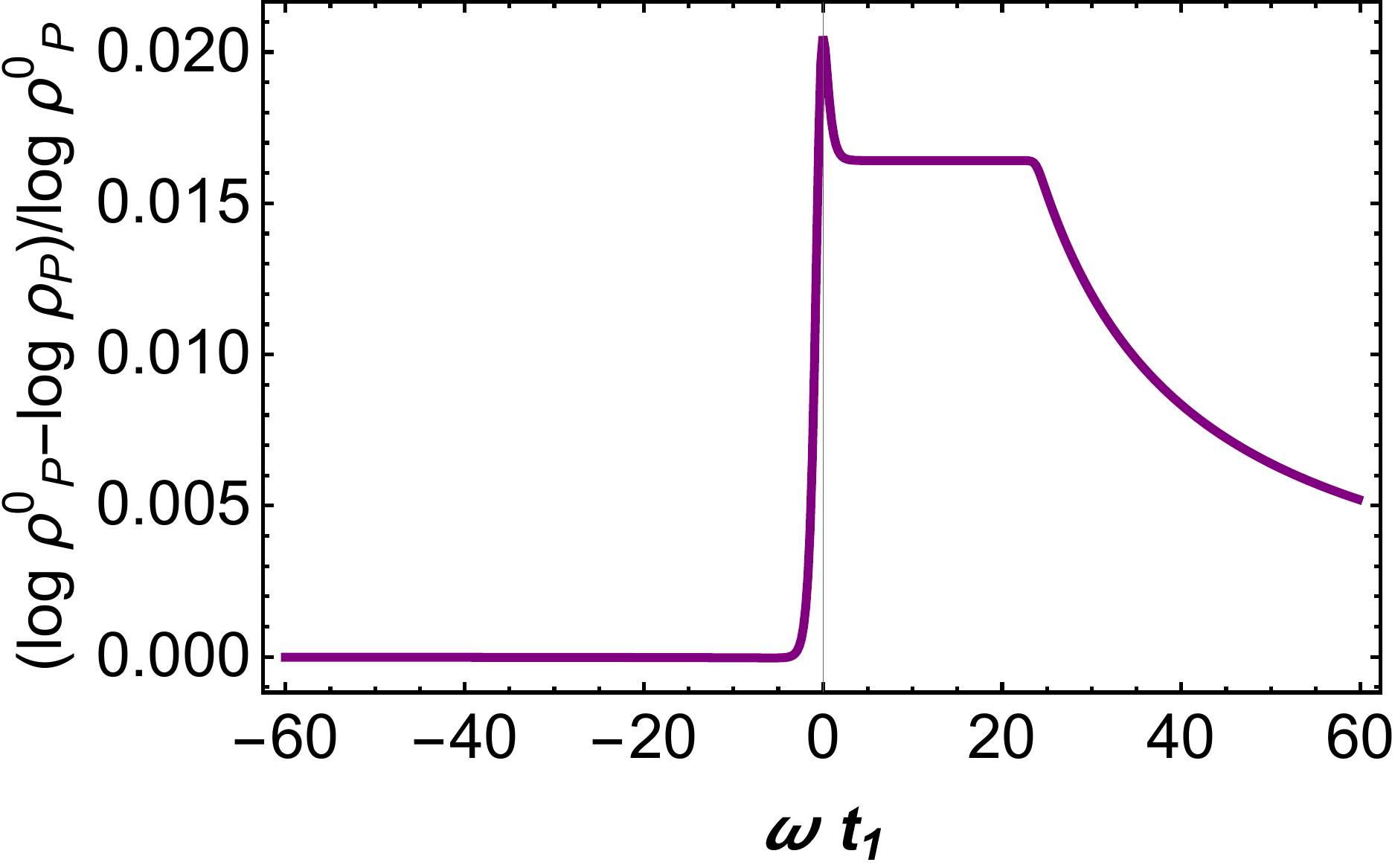}
\caption{$
\log\rho_\mt{P}$ and $\log\rho_\mt{P}^0$ vs time for the figure on the left. Relative error $(\log\rho_\mt{P}^0-\log\rho_\mt{P})/\log\rho_\mt{P}^0$ vs time for the figure on the right. The parameters are set as $\omega t_s=\omega t_f=20,{\delta\omega}/{\omega}=10^{-3}$.}
\label{F4}
\end{figure}
Comparison between the precise values of $\log\rho_\mt{P}$ and $\log\rho_\mt{P}^0$ is shown in Fig.~\ref{F4}. Note that when $t_s=t_f=0$, the Loschmidt state and precursor state have the same complexity at leading order, \ie $\rho_\mt{L}^0=\rho_\mt{P}^0$. Further, the difference in the leading order of the covariance matrix of the Loschmidt state and precursor state is given by
\begin{equation}
G_{\mt{L}}-G_{\mt{P}}\approx\left(\begin{array}{cc}
\frac{g^2}{m\omega}(1-\cosh 2 t_ 1 \delta \omega ) &\sinh 2 t_ 1 \delta \omega\\
\sinh 2 t_ 1 \delta \omega & \frac{m\omega}{g^2}(1-\cosh 2 t_ 1 \delta \omega )  \\
\end{array}\right)\approx 0.
\end{equation}
Since the time scale we examine is much smaller than the time scale corresponding to the perturbation $\frac{1}{\delta\omega}$, the two states above are identical in the leading order. This shows that a single perturbation cannot distinguish between the Loschmidt state and the precursor state.

\subsection{Double perturbation}
\begin{figure}[t!]
\centering
\includegraphics[width=3in]{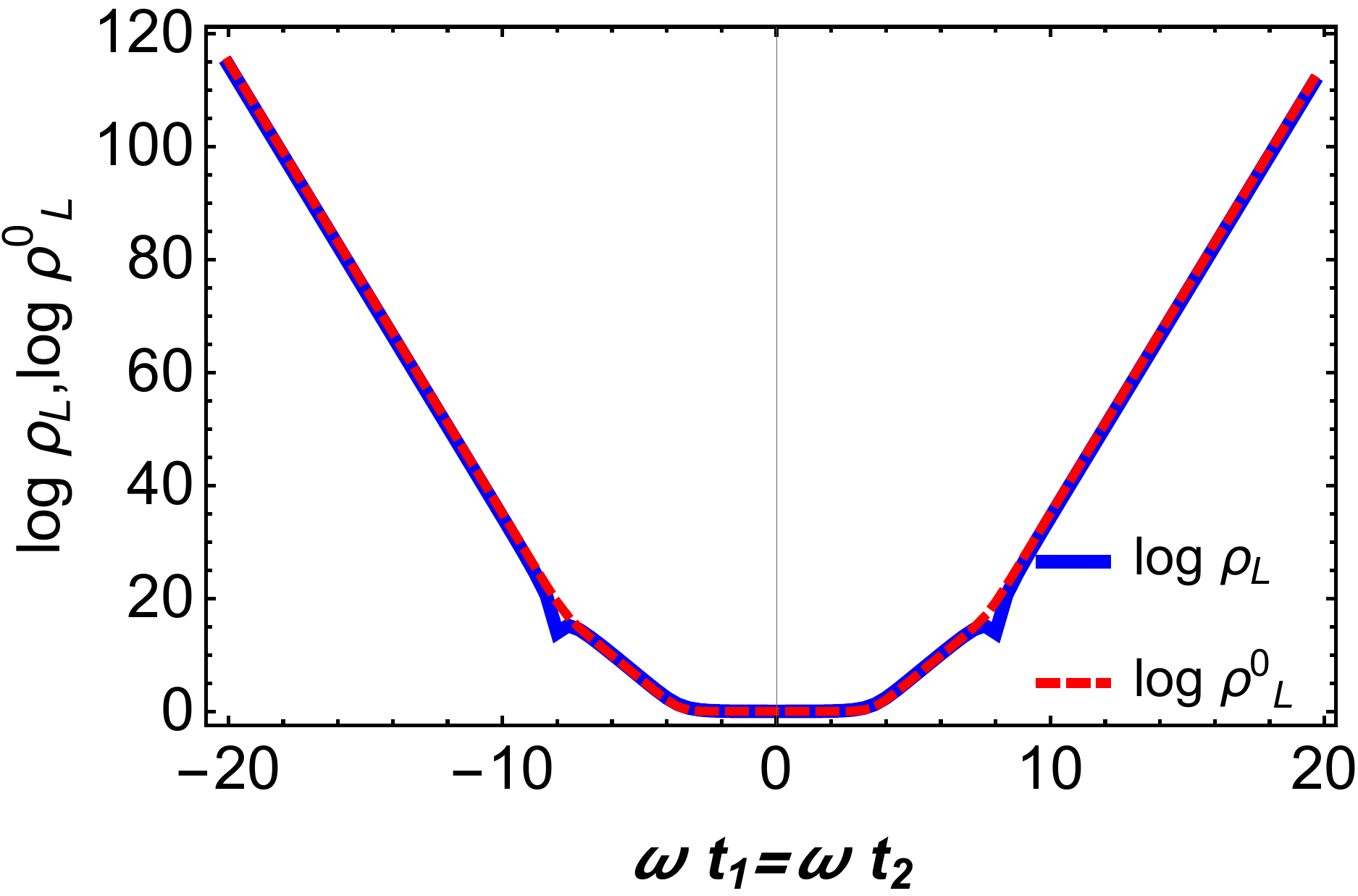}
\includegraphics[width=3in]{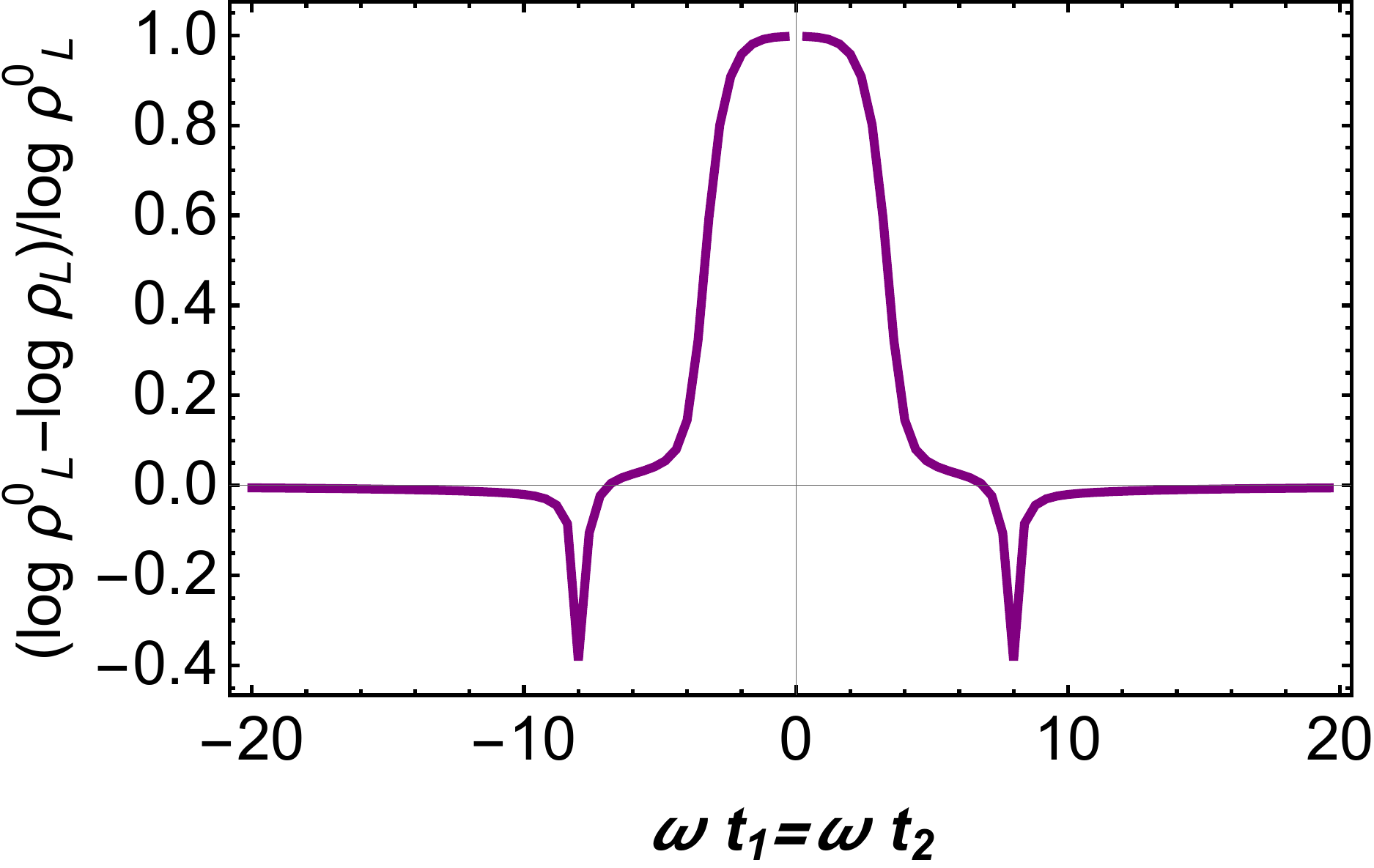}
\caption{$\log\rho_\mt{L}$ and $\log\rho_\mt{L}^0$ vs time for the figure on the left. Relative error $(\log\rho_\mt{L}^0-\log\rho_\mt{L})/\log\rho_\mt{L}^0$ vs time for the figure on the right. The parameters is set as ${\delta\omega}/{\omega}=10^{-3}$.}
\label{F5}
\end{figure}

We have looked at a single perturbation in the previous part. In this section, we look at the situation of a double perturbation, i.e., $N=2$. Similar to the previous section, the covariance matrix of the Loschmidt state is given by
\begin{equation}
G_{\mt{L}}=M'(t_2)M(t_2)M'(t_1)M(t_1)G_{\mt{R}}M(t_1)^TM'(t_1)^TM(t_2)^TM'(t_2)^T.
\end{equation}
The eigenvalue $\rho_\mt{L}$ of the relative covariance matrix $\Delta_{\mt{L}}=G_LG_R^{-1}$ at leading order is given by
\begin{equation}
\label{e36}
\begin{aligned}
\rho_\mt{L} \approx \rho_\mt{L}^0&=1
%+\frac{\delta \omega^2}{2 \omega^2}\cosh\Big[4 (t_1-\frac{t_f}{2}) \omega\Big]
%+\frac{\delta \omega^2}{2 \omega^2}\cosh\Big[4(t_2-\frac{t_f}{2})\omega\Big]\\
%&+\frac{\delta \omega^4}{8 \omega^4}\cosh\Big[4(t_1-t_2+\frac{t_f}{2}) \omega\Big]
%+\frac{\delta \omega^4}{8 \omega^4}\cosh\Big[4 (t_1+\frac{t_f}{2}) \omega\Big] \\
%&+\frac{\delta \omega^6}{32 \omega^6}\cosh\Big[4(t_1+t_2-\frac{t_f}{2}) \omega\Big]
%+\frac{\delta \omega^4}{8 \omega^4}\cosh\Big[4(t_2+\frac{t_f}{2})\omega\Big]\\
%&+\frac{\delta \omega^6}{32 \omega^6}\cosh\Big[4(t_1-t_2-\frac{t_f}{2}) \omega\Big]
%+\frac{\delta \omega^8}{128 \omega^8}\cosh\Big[4(t_1+t_2+\frac{t_f}{2}) \omega\Big],
+\alpha e^{|4  \omega t_ 1 |}
+\alpha e^{|4t_2\omega|}
+\alpha^{2} e^{|4(t_1-t_2) \omega|}
+\alpha^{3} e^{|4(t_1+t_2) \omega|},
\end{aligned}
\end{equation}
which is just Eq.~\eqref{rho_L} with $N=2$. Note that Eq.~\eqref{e36} can return to Eq.~\eqref{e37} when $t_2=0$, because the extra terms are high-order infinitesimal quantities. Comparison between the precise
values of $\log\rho_\mt{L}$ and $\log\rho_\mt{L}^0$ is shown in Fig.~\ref{F5}.

\begin{figure}[t!]
\centering
\includegraphics[width=3in]{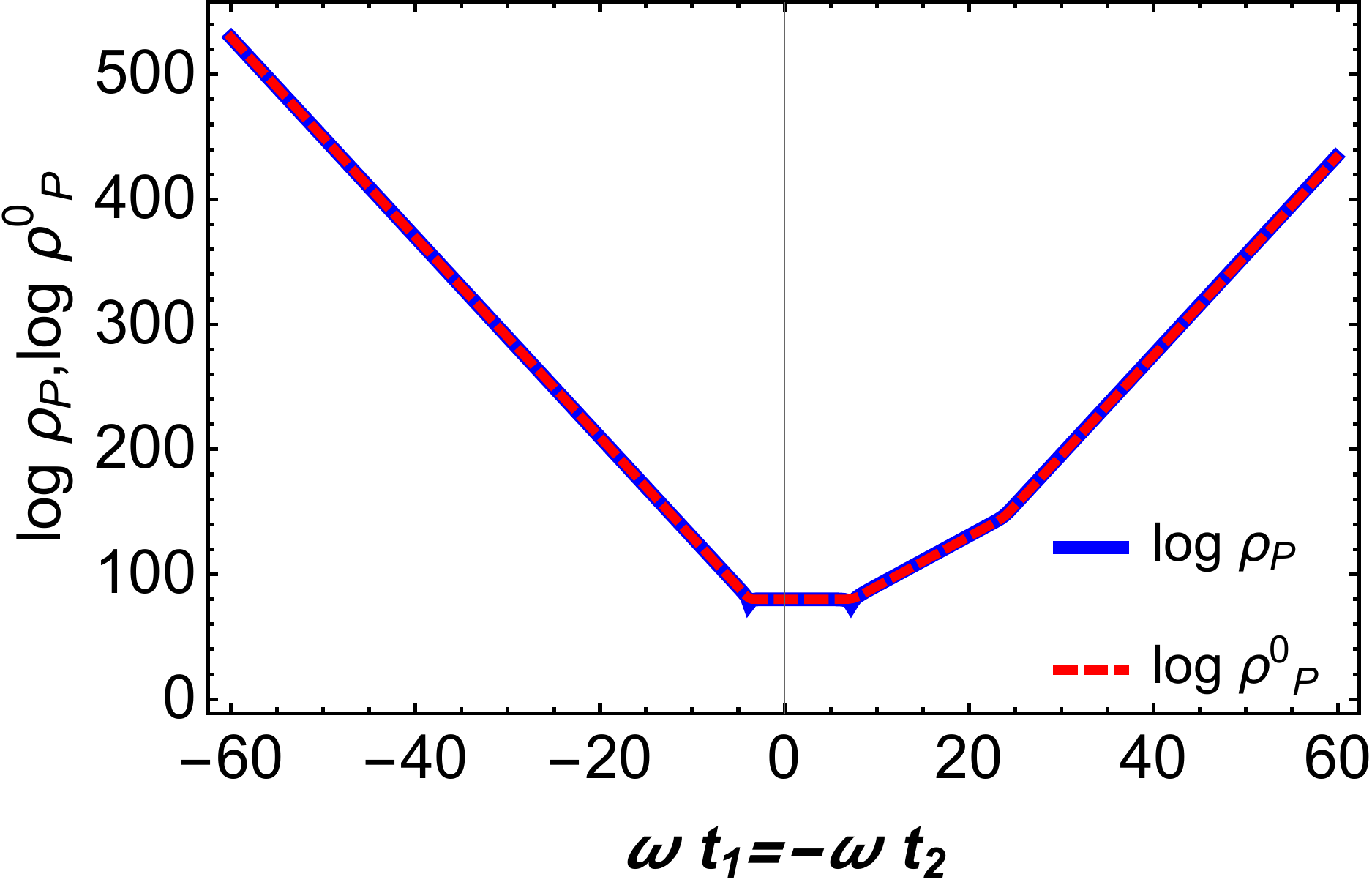}
\includegraphics[width=3in]{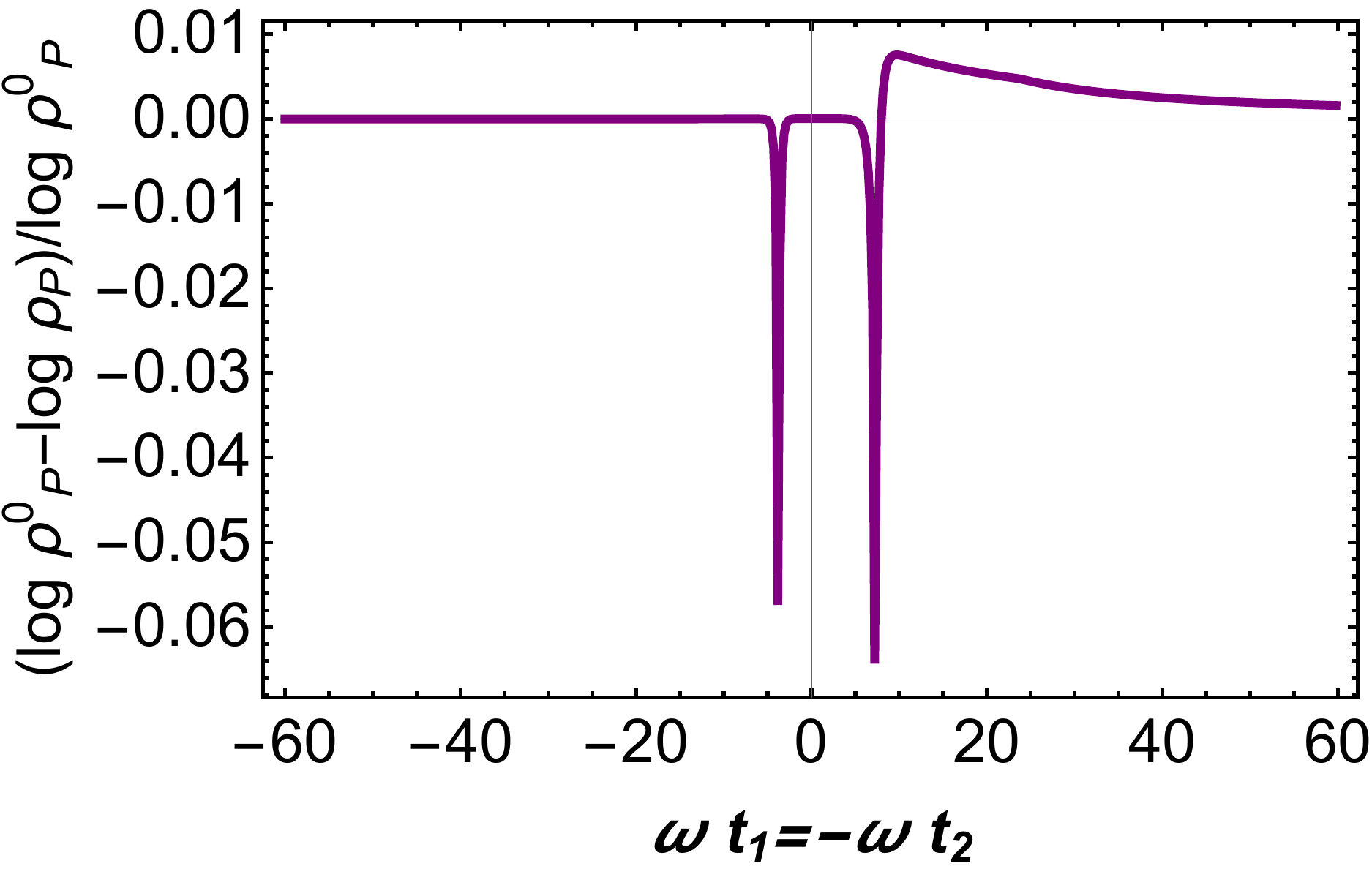}
\caption{$\log\rho_\mt{P}$ and $\log\rho_\mt{P}^0$ vs time for the figure on the left. Relative error $(\log\rho_\mt{P}^0-\log\rho_\mt{P})/\log\rho_\mt{P}^0$ vs time for the figure on the right. The parameters are set as $\omega t_s=-\omega t_f=20,{\delta\omega}/{\omega}=10^{-3}$.}
\label{F7}
\end{figure}

The covariance matrix of the precursor state is given by
\begin{equation}
\begin{aligned}
    G_{\mt{P}} &=M(t_f)M(-t_2)WM(t_2)M(-t_1)WM(t_1)M(-t_s)G_{\mt{R}}  \\
   &\times  M(-t_s)^TM(t_1)^TW^TM(-t_1)^TM(t_2)^TW^TM(-t_2)^TM(t_f)^T.\end{aligned}
\end{equation}
The eigenvalue $\rho_\mt{P}$ of the relative covariance matrix $\Delta_{\mt{P}}=G_PG_R^{-1}$ at leading order is given by
\begin{equation}
\begin{aligned}
\rho_\mt{P}  \approx  \rho_\mt{P}^0 & = e^ {|2(t_s-t_f)\omega |}
      +\alpha e^{\big|4 (\frac{t_s}{2}-t_1+\frac{t_f}{2}) \omega\big|}  \\
   &+ \alpha e^{\big|4(\frac{t_s}{2}-t_2+\frac{t_f}{2})\omega\big|}
      +\alpha^{2} e^{\big|4(\frac{t_s}{2}-t_1+t_2-\frac{t_f}{2}) \omega\big|}.
      \label{e38}
\end{aligned}
\end{equation}
The above formula is exactly Eq.~\eqref{rho_P} with $N=2$. Since we only keep terms with a growth rate of $4\omega$, terms like $\alpha \exp|2(t_s-t_1-t_2+t_f)|$ are discarded as discussed in Eq.~\eqref{e315}. Note that it appears in the leading order when $t_1=t_2$, but only contributes a term $\log2$ after taking the logarithm. So it is reasonable to ignore it. We will also use this in the following calculations and will not explain this anymore. Different from the above, Eq.~\eqref{e38} can not return to Eq.~\eqref{e39} when $t_2=0$, since the fixed perturbation $\hat{W}$ corresponding to $t_2$ does not disappear. Comparison between the precise values of $\log\rho_\mt{P}$ and $\log\rho_\mt{P}^0$ is shown in Fig.~\ref{F7}.

\subsection{Quadruple perturbation}
\begin{figure}[t!]
\centering
\includegraphics[width=3in]{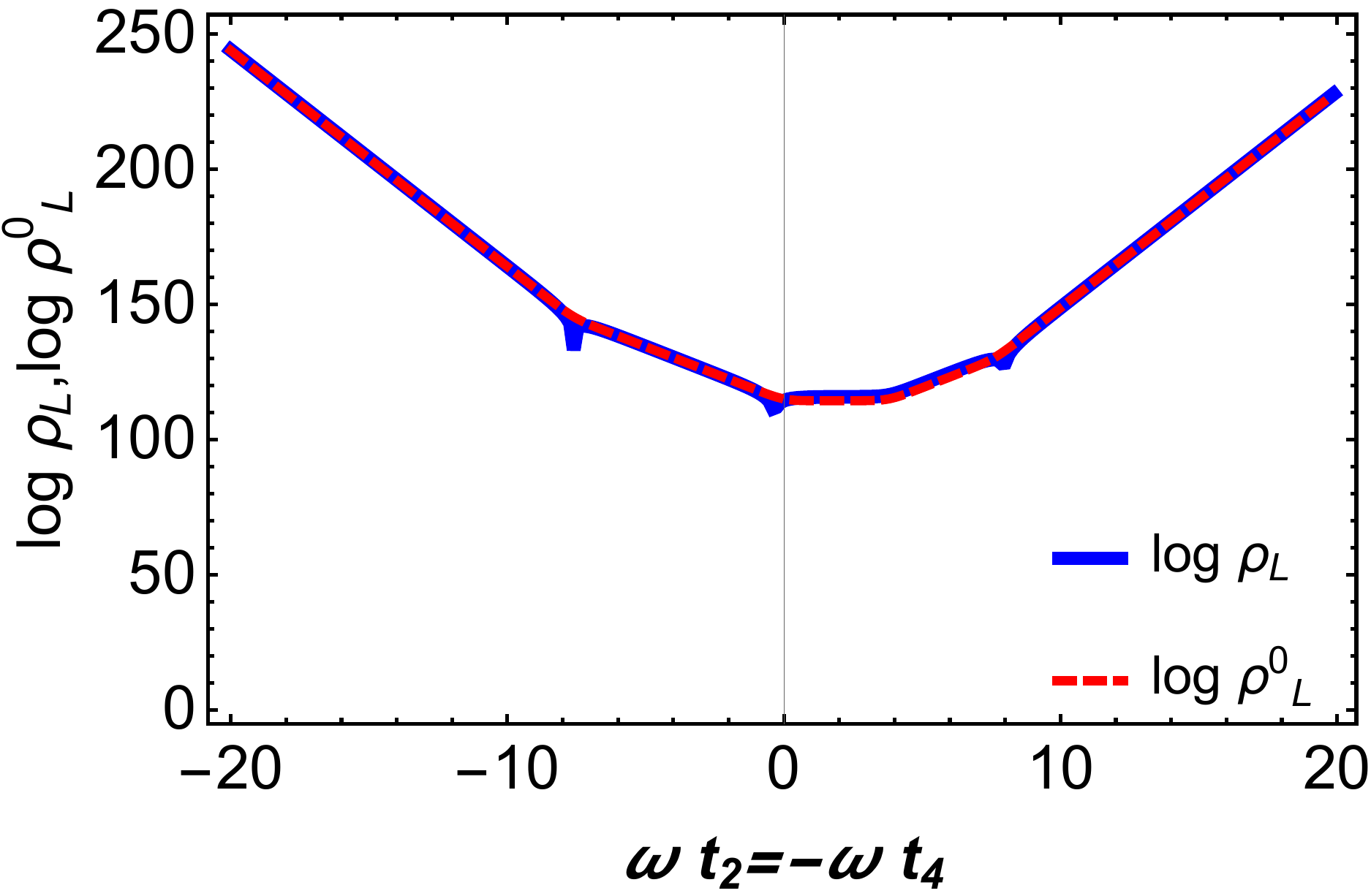}
\includegraphics[width=3in]{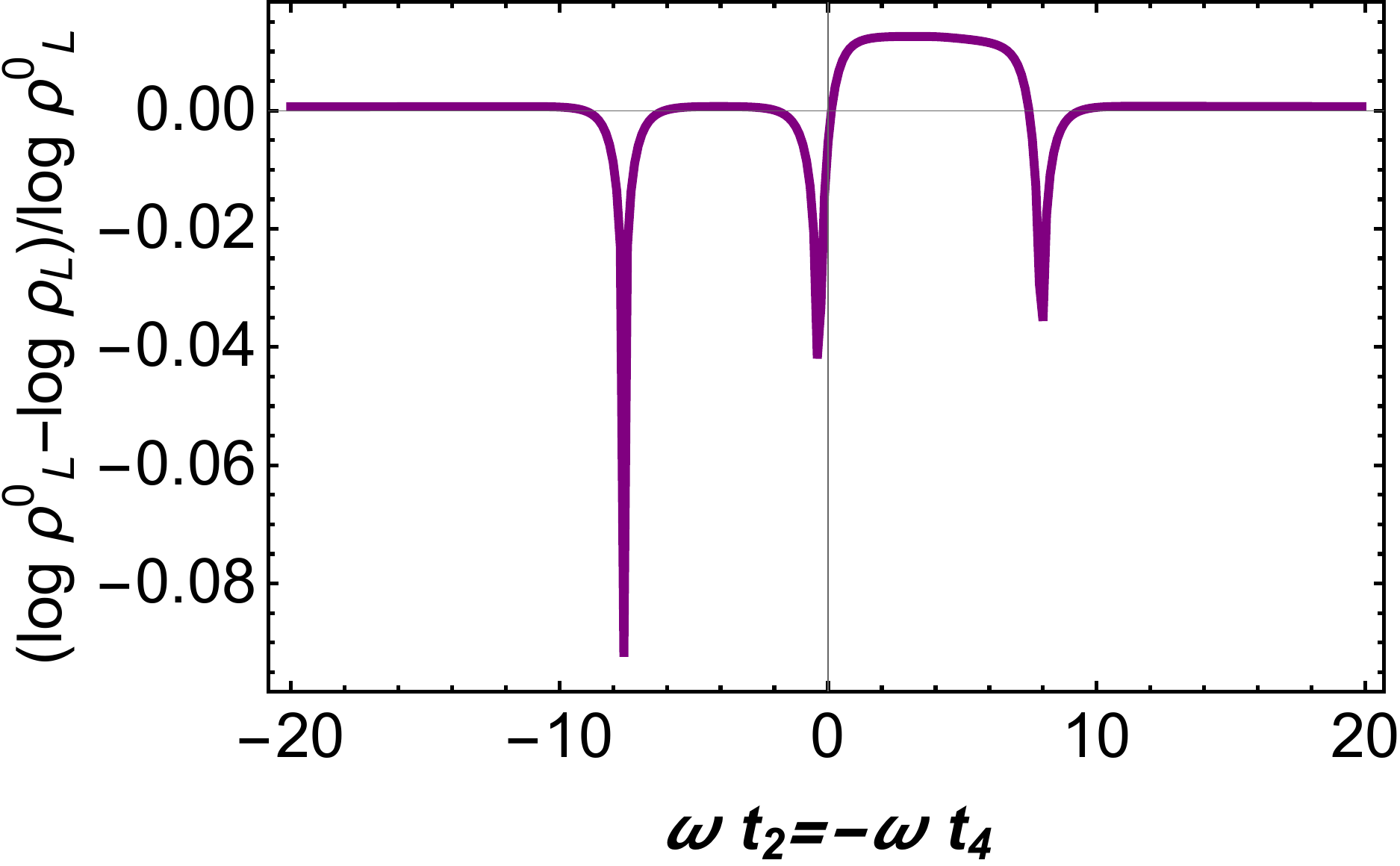}
\caption{$\log\rho_\mt{L}$ and $\log\rho_\mt{L}^0$ vs time for the figure on the left. Relative error $(\log\rho_\mt{L}^0-\log\rho_\mt{L})/\log\rho_\mt{L}^0$ vs time for the figure on the right. The parameters are set as $\omega t_1=\omega t_3=20,\frac{\delta\omega}{\omega}=10^{-3}$.}
\label{F8}
\end{figure}

In this subsection, we skip triple perturbation and discuss quadruple perturbation in order to better explain Eq.~\eqref{rho_L}, namely $N=4$. Similar to the previous subsection, the covariance matrix of the Loschmidt state is given by
\begin{equation}
\begin{aligned}
G_{\mt{L}}&=M'(t_4)M(t_4)M'(t_3)M(t_3)M'(t_2)M(t_2)M'(t_1)M(t_1)G_{\mt{R}}\\
& \times M(t_1)^TM'(t_1)^TM(t_2)^TM'(t_2)^TM(t_3)^TM'(t_3)^TM(t_4)^TM'(t_4)^T.
\end{aligned}
\end{equation}
Since the eigenvalue $\rho_\mt{L}$ of the relative covariance matrix $\Delta_{\mt{L}}=G_LG_R^{-1}$ at leading order is very complex, we only list representative parts of it
\begin{equation}
\label{e40}
        \begin{aligned}
        \rho_\mt{L} \approx \rho_\mt{L}^0 
  &=
 1
 %+\frac{\delta \omega^2 \cosh[4 (t_{1}-\frac{t_f}{2}) \omega]}{2 \omega^2}
  %      +\frac{\delta \omega^4 \cosh[4 (t_{1}-t_{2}+\frac{t_f}{2}) \omega]}{8 \omega^4}
   %     +\frac{\delta \omega^6 \cosh[4 (t_{1}+t_{2}-\frac{t_f}{2}) \omega]}{32 \omega^6}
    %    \\
     %   +&\frac{\delta \omega^6 \cosh[4 (t_{1}-t_{2}+t_{3}-\frac{t_f}{2}) \omega]}{32 \omega^6}
      % +\frac{\delta \omega^8 \cosh[4 (t_{1}+t_{2}-t_{3}+\frac{t_f}{2}) \omega]}{128 \omega^8}
      %  \\
      %  +&\frac{\delta \omega^8 \cosh[4 (t_{1}-t_{2}-t_{3}+\frac{t_f}{2}) \omega]}{128 \omega^8}
      %+\frac{\delta \omega^{10} \cosh[4 (t_{1}+t_{2}+t_{3}-\frac{t_f}{2}) \omega]}{512 \omega^{10}}
      %  \\
      %  +&\frac{\delta \omega^{8} \cosh[4 (t_{1}-t_{2}+t_{3}-t_{4}+\frac{t_f}{2}) \omega]}{128 \omega^{8}}
      %  +\frac{\delta \omega^{14} \cosh[4 (t_{1}+t_{2}+t_{3}+t_{4}-\frac{t_f}{2}) \omega]}{8192 \omega^{14}}
      %  \\
        +\alpha e^{|4 t_{1} \omega|}
        +\alpha^{2}e^{|4 (t_{1}-t_{2}) \omega|}
        +\alpha^{3}e^{|4 (t_{1}+t_{2}) \omega|}
        +\alpha^{3} e^{|4 (t_{1}-t_{2}+t_{3}) \omega|}
        \\
        &
       +\alpha^{4} e^{|4 (t_{1}+t_{2}-t_{3}) \omega|}
       +\alpha^{4}e^{|4 (t_{1}-t_{2}-t_{3}) \omega|}
       +\alpha^{4}e^{|4 (t_{1}-t_{2}+t_{3}-t_{4}) \omega|}
        \\
        &+\alpha^{5}e^{|4 (t_{1}+t_{2}+t_{3}) \omega|}
        +\alpha^{7}e^{|4 (t_{1}+t_{2}+t_{3}+t_{4}) \omega|}
        \\&+(\text{Other similar terms}),
        \end{aligned}
    \end{equation}
which is the case of Eq.~\eqref{rho_L} with $N=4$. Note that Eq.~\eqref{e40} can return to Eq.~\eqref{e36} when $t_3=t_4=0$, because the extra terms are high-order infinitesimal quantities. The physics happening here can be seen as late-time growth strengthens higher-order perturbation terms. Comparison between $\log\rho_\mt{L}$ and $\log\rho_\mt{L}^0$ is shown in Fig.~\ref{F8}.

The covariance matrix of the precursor state reads as
\begin{equation}
\begin{aligned}
G_{\mt{P}}&=M(t_f)M(-t_4)WM(t_4)M(-t_3)WM(t_3)M(-t_2)WM(t_2)M(-t_1)WM(t_1)\\ &\times M(-t_s)G_{\mt{R}}M(-t_s)^TM(t_1)^TW^TM(-t_1)^TM(t_2)^TW^TM(-t_2)^T\\ &\times M(t_3)^TW^TM(-t_3)^TM(t_4)^TW^TM(-t_4)^TM(t_f)^T.
\end{aligned}
\end{equation}
Due to the complexity of the eigenvalue of the relative covariance matrix at leading order, only sample portions are listed, \ie
\begin{equation}
\label{e41}
        \begin{aligned}
        \rho_\mt{P} \approx \rho_\mt{P}^0
 &=
 e^ {|2 (t_s-t_f) \omega|}
 +\alpha e^{\big|4 (\frac{t_s}{2}-t_{1}+\frac{t_f}{2}) \omega\big|}
 +\alpha^{2} e^{\big| 4 (\frac{t_s}{2}-t_{1}+t_{2}-\frac{t_f}{2}) \omega\big|}
       \\&+\alpha^{3} e^{\big| 4 (\frac{t_s}{2}-t_{1}+t_{2}-t_{3}+\frac{t_f}{2}) \omega\big|}
        +\alpha^{4} e^{\big| 4 (\frac{t_s}{2}-t_{1}+t_{2}-t_{3}+t_{4}-\frac{t_f}{2}) \omega\big|}
        \\&+(\text{Other similar terms}).
        \end{aligned}
    \end{equation}
The above formula is exactly Eq.~\eqref{rho_P} with $N=4$.
\begin{figure}[t!]
\centering
\includegraphics[width=3in]{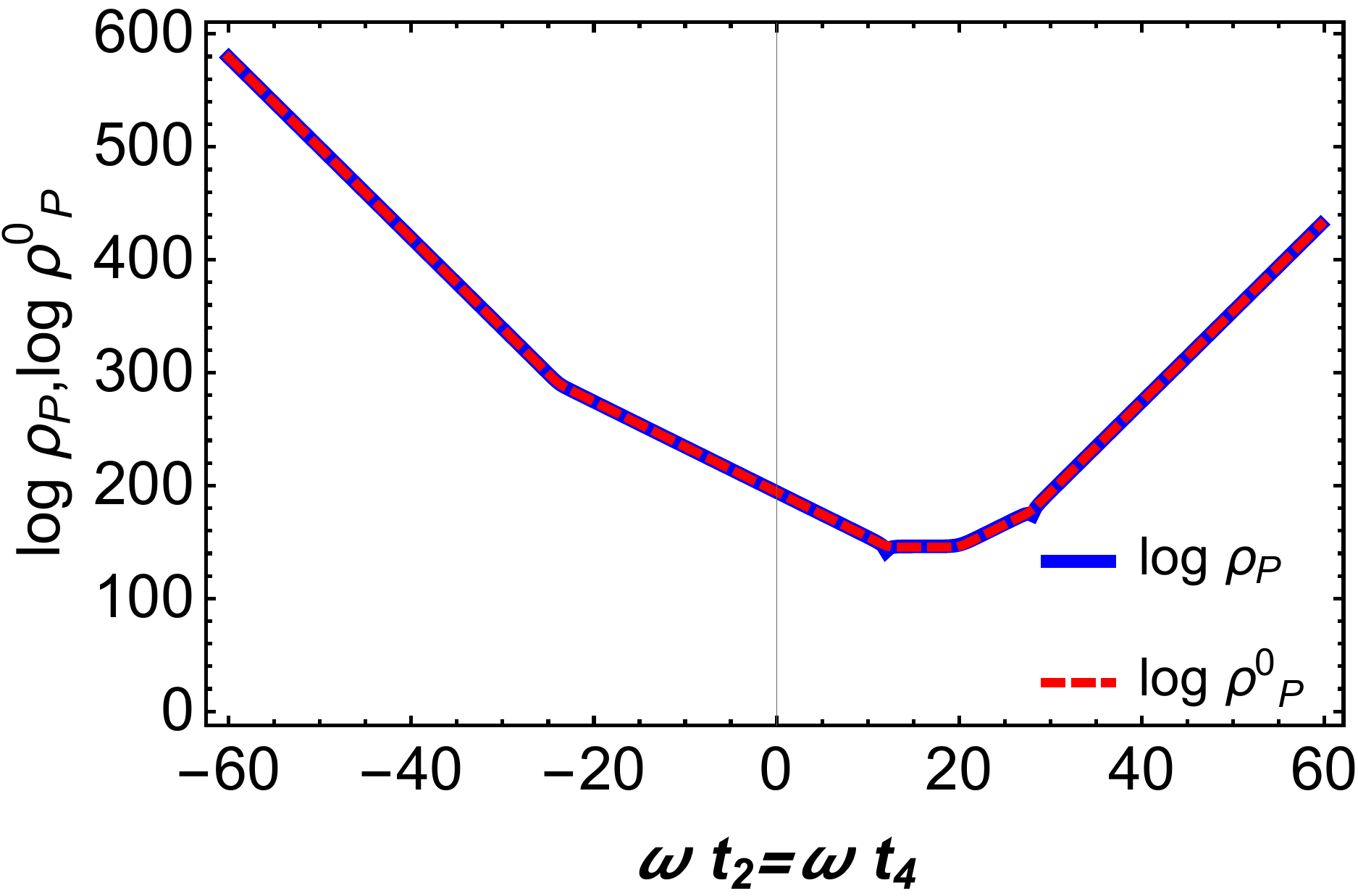}
\includegraphics[width=3in]{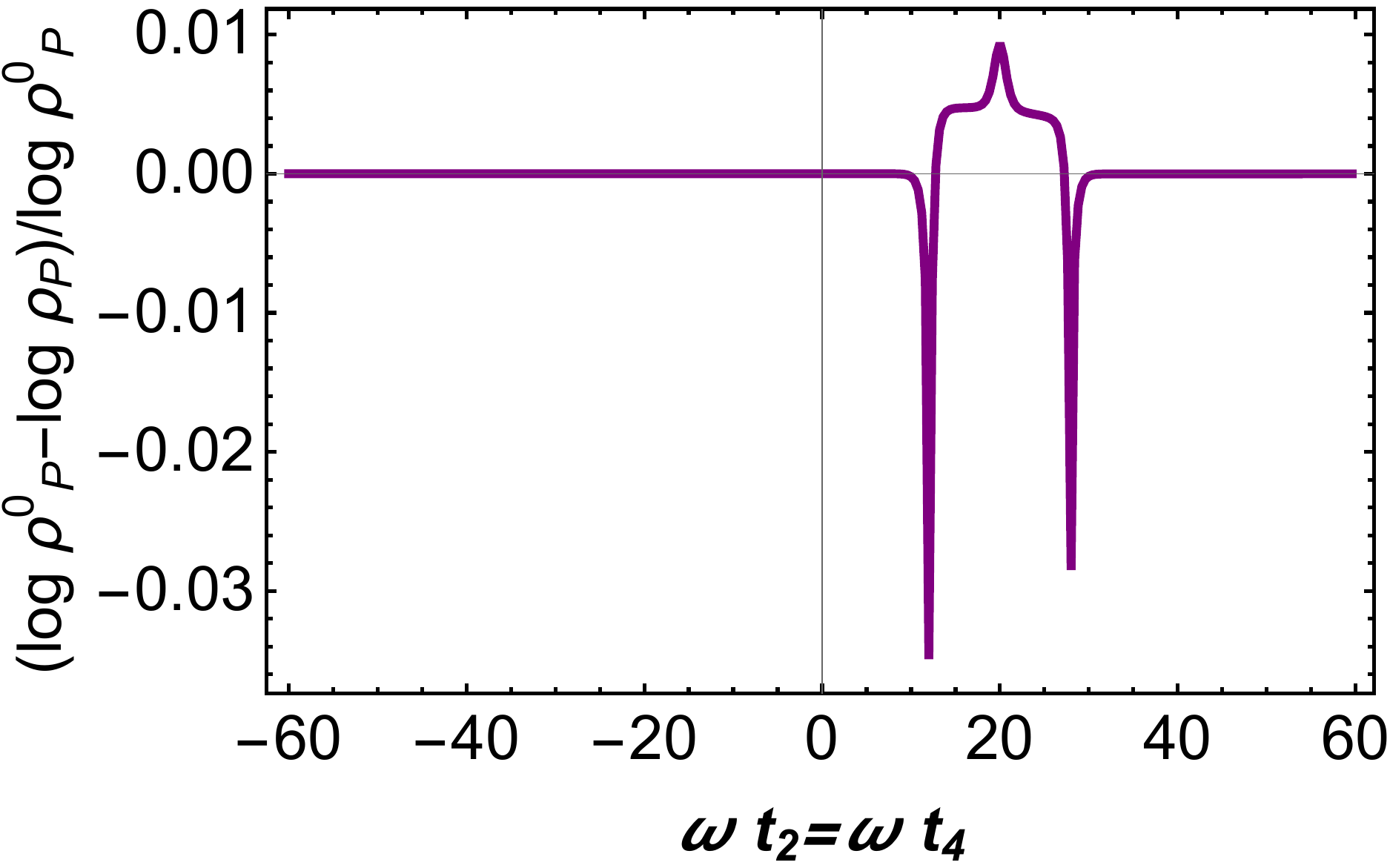}
\caption{$\log\rho_\mt{P}$ and $\log\rho_\mt{P}^0$ vs time for the figure on the left. Relative error $(\log\rho_\mt{P}^0-\log\rho_\mt{P})/\log\rho_\mt{P}^0$ vs time for the figure on the right. The parameters are set as $\omega t_1=\omega t_3=-\omega t_s=-\omega t_f=20,{\delta\omega}/{\omega}=10^{-3}$.}
\label{F9}
\end{figure}
Note that Eq.~\eqref{e41} can not return to Eq.~\eqref{e38} when $t_3=t_4=0$, because fixed perturbations corresponding to $t_3,t_4$ persist. Comparison between $\log\rho_\mt{P}$ and $\log\rho_\mt{P}^0$ is shown in Fig.~\ref{F9}.

\section{Summary}
In this article, we investigated the multifold behaviors of the inverted harmonic oscillator through the multifold complexity and Loschmidt echo. Using the covariance matrix associated with Gaussian states, we obtained the analytical expressions of the multifold complexity and Loschmidt echo. We found that the leading contributions of the multifold complexity $\mathcal{C}$ and Loschmidt echo $\mathcal{I}$ have the same expression of $\frac{1}{2}\log\rho$, where $\rho$ is give by
\begin{equation}
\begin{aligned}
\rho_\mt{L} \approx \rho_\mt{L}^0 =1+ \sum_{n=1}^{N} ~\sum_{j_1<j_2<\cdots<j_n}~ \sum_{\oplus_1,\cdots,\oplus_{n-1}}
    \alpha^{2n-1-\kappa(\oplus_1,\cdots,\oplus_{n-1})}
    e^{\left|4 (t_{j_1}\oplus_1 t_{j_2}\oplus_2  \cdots\oplus_{n-1} t_{j_n}
                 ) \omega\right|}
               ,
\end{aligned}
\end{equation}
for the Loschmidt state $|\psi_{\mt{L}}\rangle$ and
\begin{equation}
\begin{aligned}
\rho_\mt{P} &\approx \rho_\mt{P}^0 = e^ {|2 (t_s-t_f) \omega|}
   + \sum_{n=1}^{N} \sum_{j_1<j_2<\cdots<j_n}
   \alpha^{n}
     e^{\left|4 \big( {t_s}/{2}+ \sum_{k=1}^{n} (-1)^{k}t_{j_k} +(-1)^{n+1}{t_f}/{2}\big) \omega
            \right|}.
\end{aligned}
\end{equation}
for the precursor state $\left|\psi_{\mt{P}}\right\rangle$. Note that both the Loschmidt state and the precursor state exhibit the anticipated structure as deduced from Eq.~\eqref{e43} as well as more complicated aspects that may be derived from the equation above. We discovered that the latter is the combination of the terms in the first with the quickest scrambling time. In addition, we demonstrate that the longest permutation of the given time combination in an alternating zig-zag order dominates complexity, achieving the same result as Stanford and Susskind foresaw.

In the context of AdS/CFT duality, it would be interesting to investigate the above behaviors of complexity in holographic systems such as Sachdev-Ye-Kitaev model \cite{Sachdev:1992fk}. Since the spatial dimension of this model is zero, the locality of the precursor operator is manifested as infinitely close to the unit operator, which is similar to the inverted harmonic oscillator. We believe that the behavior in the leading order will be similar to the formula above.

\section*{Acknowledgments}
We would like to thank Juan F. Pedraza and Shan-Ming Ruan for their very useful suggestions, comments, and sincere help. We also thank Jing Chen and Meng-Ting Wang for their helpful discussions. This work was supported by the National Natural Science Foundation of China (Grants No. 11875151, No. 12047501), the 111 Project (Grant No. B20063), and Lanzhou City's scientific research funding subsidy to Lanzhou University.

\end{document}